\RequirePackage{fix-cm}
\documentclass[twocolumn]{svjour3}
\smartqed

\usepackage[utf8]{inputenc}
\usepackage{tikz}
\usetikzlibrary{arrows}
\usepackage{amsmath} 
\usepackage{graphicx} 
\usepackage{multirow} 
\usepackage{tabularx} 
\usepackage{color} 
\usepackage{amssymb} 
\usepackage{amsfonts} 
\usepackage{multirow}
\usepackage{booktabs}
\usepackage{graphicx}
\usepackage{pgfplots}
\usepackage{tikz}
\usetikzlibrary{patterns}
\usepackage{rotating}
\usepackage{threeparttable}
\usepackage{enumerate} 
\usepackage{comment}
\usepackage{array}
\newcolumntype{P}[1]{>{\raggedright\arraybackslash}p{#1}}
\usepackage{wrapfig}
\usepackage{afterpage}
\usepackage{todonotes}
\usepackage{booktabs,rotating,pdflscape,graphicx,listings}
\usepackage{graphicx}
\definecolor{color-1}{rgb}{0.59,0.86,1}
\definecolor{color-2}{rgb}{1,1,1}
\definecolor{color-3}{rgb}{0.27,0.45,0.77}
\definecolor{color-4}{rgb}{0.75,0.75,0.75}
\definecolor{color-5}{rgb}{0.77,0.93,1}
\definecolor{color-6}{rgb}{0.2,0.2,0.2}
\definecolor{color-7}{rgb}{0.61,0,0.02}
\usepackage[hyphens]{url}
\usepackage[hidelinks]{hyperref}

\begin{document}

\title{Citation Recommendation: Approaches and Datasets}

\author{Michael F{\"a}rber \and Adam Jatowt}

\institute{Karlsruhe Institute of Technology (KIT),
              Kaiserstr. 89, 
              76133 Karlsruhe, Germany,  
              \email{michael.faerber@kit.edu}           
\and Kyoto University, 
              Yoshida-Honmachi, Sakyo-ku, Kyoto 606-8501, Japan,
              \email{adam@dl.kuis.kyoto-u.ac.jp}  
}

\date{Received: June 2019 / Accepted: April 2020}

\maketitle

\begin{abstract}
Citation recommendation describes the task of recommending citations for a 
given text. Due to the overload of published scientific works in recent years on the one hand, and the need to cite the most appropriate publications when writing scientific texts on the other hand, citation recommendation has emerged as an important research topic. In recent years, several approaches and evaluation data sets have been presented. However, to the best of our knowledge, no literature survey has been conducted explicitly on citation recommendation. In this article, we give a thorough introduction into automatic citation recommendation research. We then present an overview of the approaches and data sets for citation recommendation and identify differences and commonalities using various dimensions. Last but not least, we shed light on the evaluation methods, and outline general challenges in the evaluation and how to meet them. We restrict ourselves to citation recommendation for scientific publications, as this document type has been studied the most in this area. However, many of the observations and discussions included in this survey are also applicable to other types of text, such as news articles and encyclopedic articles. 
\end{abstract}

\section{Introduction}
\label{sec:introduction} 

\begin{figure}[tb]
 \centering
 \includegraphics[width=\linewidth]{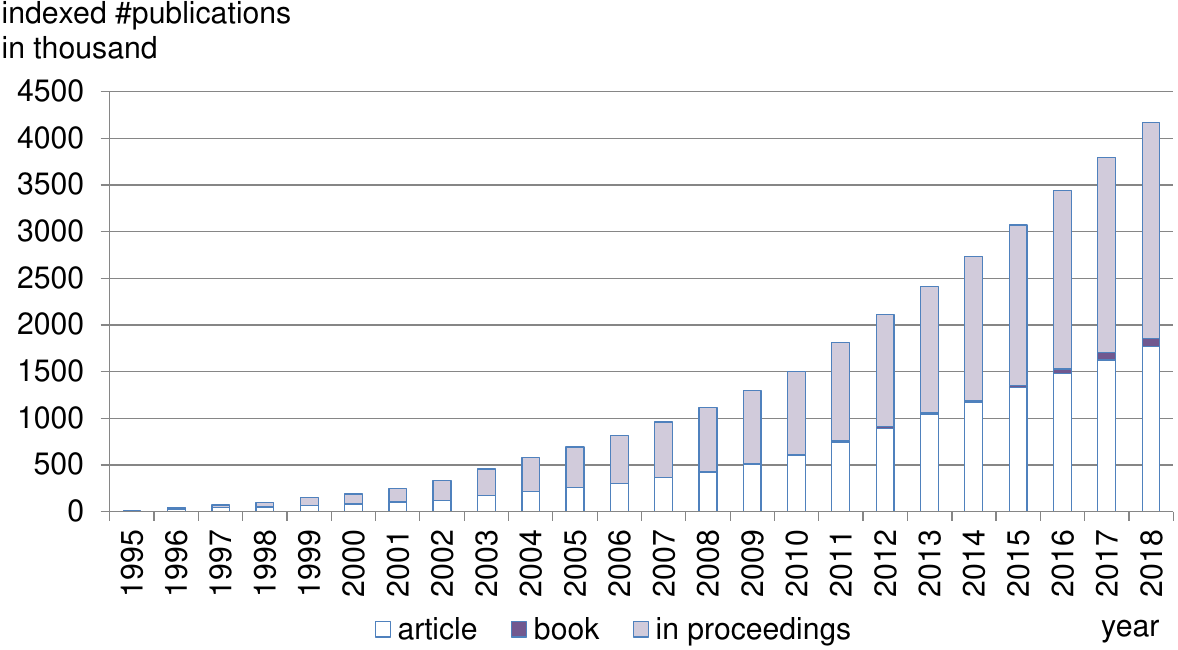}
 \caption{Growth of scientific publications indexed in DBLP from 1995 until 2018. Data source: \url{http://dblp.org/statistics/recordsindblp}.}
 \label{fig:publication-growth}
\end{figure}

Citing sources in text is essential in many scenarios. Most prominently, citing has always been an integral part of academic research. 
Scientific works need to contain appropriate citations to other works for several reasons~\cite{Teufel2009annotation}. Most notably, all claims written by the author need to be backed up in order to ensure transparency, reliability, and truthfulness. 
Secondly, mentions of methods and data sets and further important domain-specific concepts need to be linked via references in order to help the reader to properly understand the text and to give attribution to the corresponding publications and authors (see Table~\ref{tab:example-citation-types}).
However, citing properly has become increasingly difficult due to the dramatically increasing number of scientific publications published each year~\cite{Bornmann2015,Ware2015STM,Fortunato2018science} (see also Fig.~\ref{fig:publication-growth}). 
For instance, in the computer science domain alone, more than 100,000 new papers are published every year and three times more papers were published in 2010 than in 2000~\cite{Kucuktunc2014}. 
A similar trend can be observed in other disciplines~\cite{Larsen2010}. For instance, in the medical digital library database PubMed, the number of publications in 2014 (514k) was more than triple the amount published in 1990 (137k) and more than 100 times the amount published in 1950 (4k)~\cite{PubMedStats}. 
Due to this phenomenon of information  overload in science in the form of a ``tsunami of publications,'' citing appropriate publications has become an increasing challenge for scientific writing.

As a consequence, approaches for citation recommendation have been developed.
Citation recommendation refers to the task of recommending appropriate citations for a text passage within a document. 
For instance, given the phrase ``and similarly, the emergence of GANs has led to significant improvements in human image synthesis'' within a document, a citation recommendation system might insert two citations as follows: ``and similarly, the emergence of GANs [1] has led to significant improvements in human image synthesis [2].'' This would mean adding corresponding references to (1) a publication introducing generative adversarial networks (GANs), and (2) a publication backing up the statement concerning improvements in human image synthesis. Added references in such a scenario need to fit semantically to the context within the citing document and may be required to meet further constraints (e.g., concerning their recency).

\begin{table}[tb]
\caption{Examples for in-text citations from F{\"a}rber and Sampath~\cite{Faerber2019TPDL}.}
\label{tab:example-citation-types}
\centering
\begin{small}
\begin{tabular}{lp{5.4cm}}
\toprule
Citation type & ~Example sentence \\
\midrule
\textsc{concept} & ``To this end, SWRL [14] extends OWL-DL and OWL-Lite with Horn clauses.'' \\
\textsc{claim} & ``In the traditional hypertext Web, browsing and searching are often seen as the two dominant modes of interaction (Olston \& Chi, 2003).''\\
\textsc{author} & ``Gibson et al. [12] used hyperlink for identifying communities.'' \\
\bottomrule
\end{tabular}
\end{small}
\end{table}

Note that citation recommendation differs from paper recommendation~\cite{Beel2016,Steinert17}: 
\textit{paper recommendation} aims to recommend documents to the user that are worthwhile to read and to investigate (particularly, in the context of a research topic). 
To that end, one or several papers \cite{White2016,HagenBGKS16,Ahmad2017combining,SharmaGM17} or the user's already clicked/ bookmarked/written documents \cite{AlzoghbiA0L15,LiBSC18} can, for instance, be used for the recommendation.
We can refer to ~\cite{Beel2016,PaperRecSurvey2019} for surveys on paper recommendation. 
\textit{Citation recommendation}, by contrast, assists the user in substantiating a given text passage (e.g., written claim or scientific concept) within an input document by recommending publications that can be used as citations. 
The textual phrase to be backed up can vary in length -- from one word up to a paragraph -- and is called \textit{citation context}. 
In some cases \cite{He2011WSDM,Livne2014SIGIR}, the citation context needs to be discovered before the actual citation recommendation. 
While some existing works consider citation recommendation as a task of extending the set of known references for a given paper \cite{Jia2017SNA,Jia2018ECIR,Gori2006}, we consider citation recommendation purely as a task for substantiating claims and concepts in the citation context. This makes citation recommendation context-aware and very challenging, because the concept of relevance is much stricter than in ad hoc retrieval \cite{Strohman2007SIGIR}. 
Consequently, citation recommendation approaches have been proposed using additional information besides the citation context for the recommendation, such as the author's name of the input document~\cite{Ebensu2017}. 
Evaluating a citation recommendation approach requires to verify if the recommended papers are relevant as citations for given citation contexts.  
For scalability reasons, usually the citations in existing papers and their citation contexts are used as ground truth (see Sec.~\ref{sec:evaluation-methods-citation-recommendation}).

Existing surveys focus only on related research areas of citation recommendation, but not explicitly on citation recommendation itself. 
Among the most closely related studies are the surveys on paper recommendation~\cite{Beel2016,PaperRecSurvey2019}. In these articles, the authors do not consider recommender systems for given citation contexts. 
Several surveys on other aspects of citation contexts have also been published.
Alvarez et al.~\cite{Alvarez2016} summarize and discuss works on the identification of citation contexts, on the classification of each citation's role (called \textit{citation function}), and on the classification of each citation's ``sentiment'' (called \textit{citation polarity}). 
Ding et al.~\cite{Ding2014JASIST} focus on the content-based analyses of citation contexts, while White~\cite{White2004AL} considers primarily the classification of citations into classes, the topics covered by citation contexts, and the motivation of citing. 
Moreover, distantly related to this survey, surveys on the analysis of citing behavior~\cite{Bornmann2008,Tahamtan2018} and surveys on works about the analysis of citation networks exist, for instance, for the purpose of creating better measurements of the scientific impact of researchers or communities~\cite{Todeschini2016}. 
Dedicated approaches and data sets for citation recommendation are not covered in all those works, nor is there any analysis of citation recommendation evaluations and evaluation challenges. 
This makes it necessary to consider citation recommendation separately and to use task-specific dimensions for comparing the approaches.

We make the following contributions in this survey:
\begin{enumerate}
 \item We describe the process of citation recommendation, the scenarios in which it can be applied, as well as the advantages it has in general. 
 \item We systematically compare citation recommendation to related tasks and research topics. 
 \item We outline the different approaches to citation recommendation published so far and compare them by means of specifically introduced dimensions. 
 \item We give an overview of evaluation data sets and further working data sets for citation recommendation and show their limitations. 
 \item We shed light on the evaluation methods used so far for citation recommendation, we point out the challenges of evaluating citation recommendation approaches, and present guidelines for improving citation recommendation evaluations in the future. 
 \item We outline research directions concerning citations and their recommendations.
\end{enumerate}

Several reader groups can benefit from this survey: non-experts can obtain an overview of citation recommendation; the community of citation recommendation researchers can use the survey as the basis for discussions of critical points in approaches and evaluations, as well as for getting suggestions for future research directions (e.g., research topic suggestions for PhD candidates); and finally, the survey can assist developers in choosing among the available approaches or data sets. 

The rest of this article is structured as follows: in Section~\ref{sec:task-description-related-fields}, we introduce the field of citation recommendation to the reader. 
In Section~\ref{sec:comparison-approaches}, we describe how we collected publications presenting citation recommendation approaches. We propose classification dimensions and compare the approaches by these dimensions.
In Section~\ref{sec:data-sets}, we give an overview of evaluation data sets and compare the data sets by corresponding dimensions. 
Section~\ref{sec:evaluation-methods-challenges} gives a systematic overview of the evaluation methods that have been applied so far and of the challenges that emerge when evaluating citation recommendation approaches. 
Section~\ref{sec:potential-future-work} is dedicated to potential future work. The survey closes in Section~\ref{sec:conclusions} with a summary.

\section{Citation Recommendation}
\label{sec:task-description-related-fields}

\subsection{Terminology}
\label{sec:terminology}

In the following, we define some important concepts of citation recommendation, which we use throughout the article. 
In order to have a generic task formalization, as we prefer, 
we do not restrict ourselves to scientific papers as a document type, but consider text documents in general.

\begin{figure}[tb]
 \centering
 \includegraphics[width=0.97\linewidth]{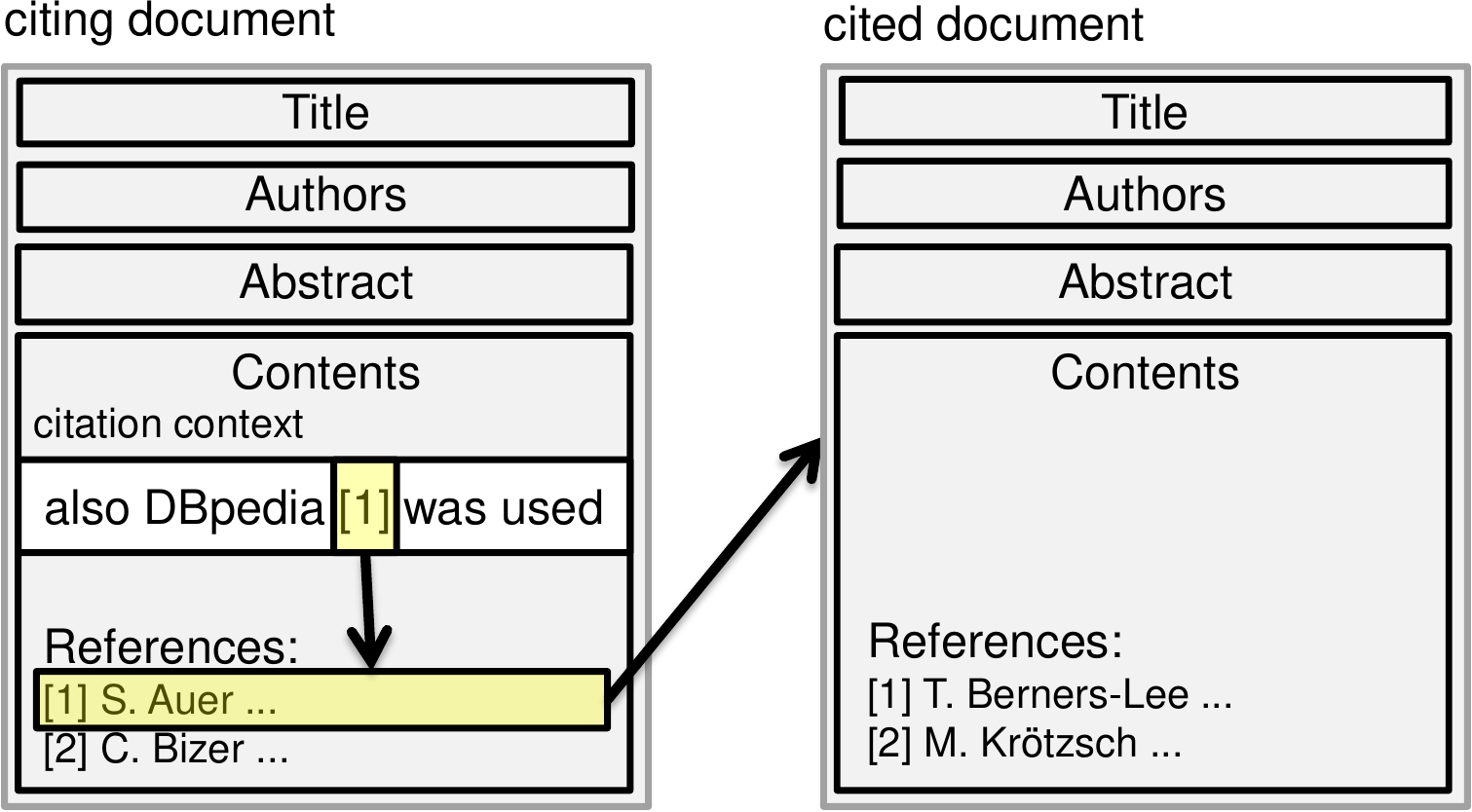}
 \caption{Visualization of a citation in a scientific paper.}
 \label{fig:terminology-figure}
 \end{figure}
 
The basic concept of citing is depicted in Fig.~\ref{fig:terminology-figure}. A \textit{citation} is defined as a link between a \textit{citing document} and a \textit{cited document} at a specific location in the citing document. This location is called the \textit{citation marker} (e.g., ``[1]'') and the text fragment which should be supported by the citation is called the \textit{citation context}. 
During processing, the citation context can be transformed into an \textit{abstract representation}, such as an embedding vector \cite{Ebensu2017,Bhagavatula2018} or a translation model \cite{Huang2012CIKM,DBLP:conf/jcdl/HuangWMG14}. 
This enables us to more accurately match the information in the citation context with the information provided in the  \textit{``citable'' documents} (also called \textit{candidate cited documents}).

``References'' and ``citations'' are often used interchangeably in the literature. However, we name in-text references, given by citation markers, \textit{citations}. \textit{References}, in contrast, are listed in the \textit{reference section} of the citing document and describe links to other documents on a document level without context. 

In the academic field, both the citing documents and the cited documents are usually scientific \textit{papers}. We use the terms \textit{paper}, \textit{publication}, and \textit{work} interchangeably in this article. 
The \textit{authors} of scientific papers are usually \textit{researchers}. We then use \textit{researcher} and \textit{scientist} interchangeably. 
Researchers who use a citation recommendation system become \textit{users}.

Citing documents and cited documents consist of \textit{content} and \textit{metadata}. In the case of scientific papers,  the paper's metadata typically consists of the title, the author information, an abstract, and other information, such as the venue in which the paper has been published.

Different citation context lengths can be used for citation recommendation. If only a fragment of an input text document is used as citation context (e.g., a sentence \cite{He2012SPIRE,Huang2012CIKM} or a window of 50 words), we call it \textit{local citation recommendation} or \textit{context-aware citation recommendation}. If no specific citation context, but instead the whole input text document or the document's abstract is used for the recommendation (see, e.g., \cite{Strohman2007SIGIR,Nallapati2008,Tang2009PAKDD,Kataria2010}), we call it \textit{global citation recommendation} or \textit{non-context-aware citation recommendation} (following He et al.~\cite{He2010WWW}). 
In the following sections, we will primarily focus on \textit{local citation recommendation}, since only this variant targets the recommendation of papers for backing up single concepts and claims in a text fragment (i.e., assists the user in the actual citing process) and has not been addressed in other surveys, to the best of our knowledge.\footnote{It should be noted that it is also possible to design global context-aware citation recommendation approaches, i.e., approaches which recommend citations for specific contexts (e.g., sentences) but which take the whole paper into account (e.g., to ensure an even greater understanding of the context or to diversify the recommendations). However, we are not aware of any such approach being published (see also Section~\ref{sec:potential-future-work} for potential future work).}

\subsection{Scenarios, Advantages, and Caveats of Citation Recommendation} 
\label{sec:scenarios-citation-recommendation}

In the ``traditional'' process of finding appropriate citations, the researcher needs to come up with candidate publications for citing on her own. The candidate papers that can be cited are either already known by her, are contained in a given document collection, or first need to be discovered. 
For the last option, the scientist typically uses widely used bibliographic databases, such as Google Scholar,\footnote{\url{http://scholar.google.com/}} or domain-specific platforms such as DBLP\footnote{\url{http://dblp.org/}} or PubMed.\footnote{\url{http://www.ncbi.nlm.nih.gov/pubmed/}} 
The search for candidate papers to cite typically requires considerable time and effort as well as skills: the right keywords for querying need to be found, and the top $n$ returned documents need to be manually assessed with regards to their relevance to the citing document and to the specific citation contexts.

The idea of citation recommendation is to enhance the citing process: The user provides the text she has written (with or without initial citations) to the recommender system. This system then presents to the user for specific segments of the input text all publications which were determined automatically as suitable citations. The user can investigate the recommendations in more detail and approve or disapprove them. Following this procedure, the tedious manual, separate search in bibliographic databases and paper collections can be considerably reduced (and maybe even skipped). The user does not need to think of meaningful keywords for searching papers any more. Last but not least, citing may become less dependent on the (often very limited) set of papers known to the current user. 

We do not want to hide that citation recommendation can also entail problematic features if applied inadequately. Firstly, if citing becomes purely automated, the role of citations might change (e.g., instead of criticizing, citations might support a statement; see \cite{Teufel2009annotation,Moravcsik1975,Teufel2006EMNLP} for citation function schemes). The trust in citations might decrease, since machines (here: recommender systems) might not engender as much trust as experts who have dealt with the topic. We thus argue that a human-in-the-loop is still needed for citation recommendation. 
Secondly, if the recommendation models are trained on a fixed publication data set, instead of removing citation biases, the recommender systems could introduce additional biases towards specific papers. Therefore, it must be ensured that a sufficiently large number of papers is indexed and that the new papers are indexed periodically. Caveats of citation recommendation are discussed in depth in Sec.~\ref{sec:evaluation-methods-challenges}. 

Citation recommendation systems can be designed for several user groups:

\textbf{(1) Expert Setting.} In this setting, a researcher is familiar with her research area and is in the process of writing an expert text, such as a scientific publication (e.g., after having developed a novel approach or for conducting a survey in her research field). 
Recommendations of citations can still be beneficial for her, as such a user might still be unaware of publications in their field in the light of the ``tsunami of publications'' common in all scientific fields nowadays \cite{Bornmann2015,Ware2015STM,Fortunato2018science}. 
Citation recommendation systems might come up with recommendations which were not in the focus of the researcher if she cited in the traditional way, since the system might be able to bridge language barriers \cite{Jiang2018JCDL,Tang2014SIGIR} and also find publications which use synonyms or otherwise related concepts. 

\textbf{(2) Non-Expert Setting.} 
We can think of several non-expert user types for which citation recommendation can be beneficial:
\begin{itemize}
 \item A researcher needs to write a scientific text on a topic that is outside of her core research area and expertise (e.g., generic research proposals \cite{Bethard2010} and potential future work descriptions). 
 \item A journalist in the science domain -- e.g., authoring texts for a popular science magazine -- needs to write an article on a certain scientific topic \cite{Ravenscroft2018,Peng2016}. We can assume that the journalist typically is not an expert on the topic she needs to write about. Having citations in the text helps to substantiate the written facts and make the text more complete and understandable. 
 \item ``Newcomers'' in science, such as Masters students and PhD students in their early years, are confronted with the vast amount of citable publications and typically do not know any or all of the relevant literature in the research field yet \cite{He2011WSDM,Yin2017APWeb}. Getting citations recommended helps not only students in writing systematic and scientific texts, such as research project proposals (exposés), but also their mentors (e.g., professors).
\end{itemize}
In all these non-expert settings, the relevance of the recommended citations is presumably not so much determined by the timeliness of the publications, as in the expert setting, but instead more by the general importance and prominence of the publications. 
Thus, the relevance function for finding the most appropriate citations might vary from setting to setting. 

Besides the pure topical relevance of recommended citations, also the fit from a social perspective might be essential. 
In recent decades, the citing behavior of scientists has been studied extensively in order to find good measurements for the scientific impact of scientists and their publications \cite{Bornmann2008}.
In this context, several biases in citing have been considered. Most notably, the hypothesis has been made that researchers tend to cite publications which they have written themselves or which have been written by colleagues~\cite{Hyland2003}. Another hypothesis is that very prominent and highly cited works get additional citations only due to their prominence and visibility in the community (see, e.g., \cite{White2004AL}). 
Citation recommendation systems can help in reducing biases by recommending citations which are the best fit for the author, the citation context with its argumentation, and the community.\footnote{However, please also note the caveats of citation recommendation as outlined above and in Sec.~\ref{sec:guidelines}.} Section~\ref{sec:evaluation-challenges} discusses citing bias in the context of citation recommendation in detail. 

Overall, we can summarize the benefits of citation recommendation as follows:
\begin{enumerate}
 \item Finding suitable citations should become more effective. This is because the match between the query (citation context) and the citable documents is more sophisticated than via manual matching (e.g., also considering synonyms, related topics, etc.). Furthermore, the recommender system typically covers a much larger collection of known publications than the set of documents known to the user.
 \item Researchers are more (time-)efficient during the process of citing, as the number and extent of manual investigations (using bibliographic databases or own document collections) are reduced, and because recommendations are returned immediately.
 \item The search for publications which can be cited becomes easier and more user-friendly (``citing for everyone''). As a consequence, citing is no longer just a ``privilege'' for experts, but potentially something for almost anyone.
 \item By establishing a formal relevance function dealing with the issue of which papers are cited and what characteristics they have, the process is no longer left to chance. Hence, biases in citing behavior can be minimized.
 \item Ideally, citation recommendation systems only recommend citations for valid statements and existing concepts, while unexaminable statements are not cited. Hence, citation recommendation implies an implicit fact checking process by showing sources to the user which support the written statements.
 \item Advanced citation recommendation systems can, in addition, search for suitable, cite-worthy publications in other languages than the citing document (cross-linguality). They can also recommend publications under the special consideration of topic evolution over time, of current buzzwords, or in a personalized way, by incorporating user profiles.
\end{enumerate}

\subsection{Task Definition}
\label{sec:formalization-citation-recommendation}

\begin{figure*}
 \centering
 \includegraphics[width=\textwidth]{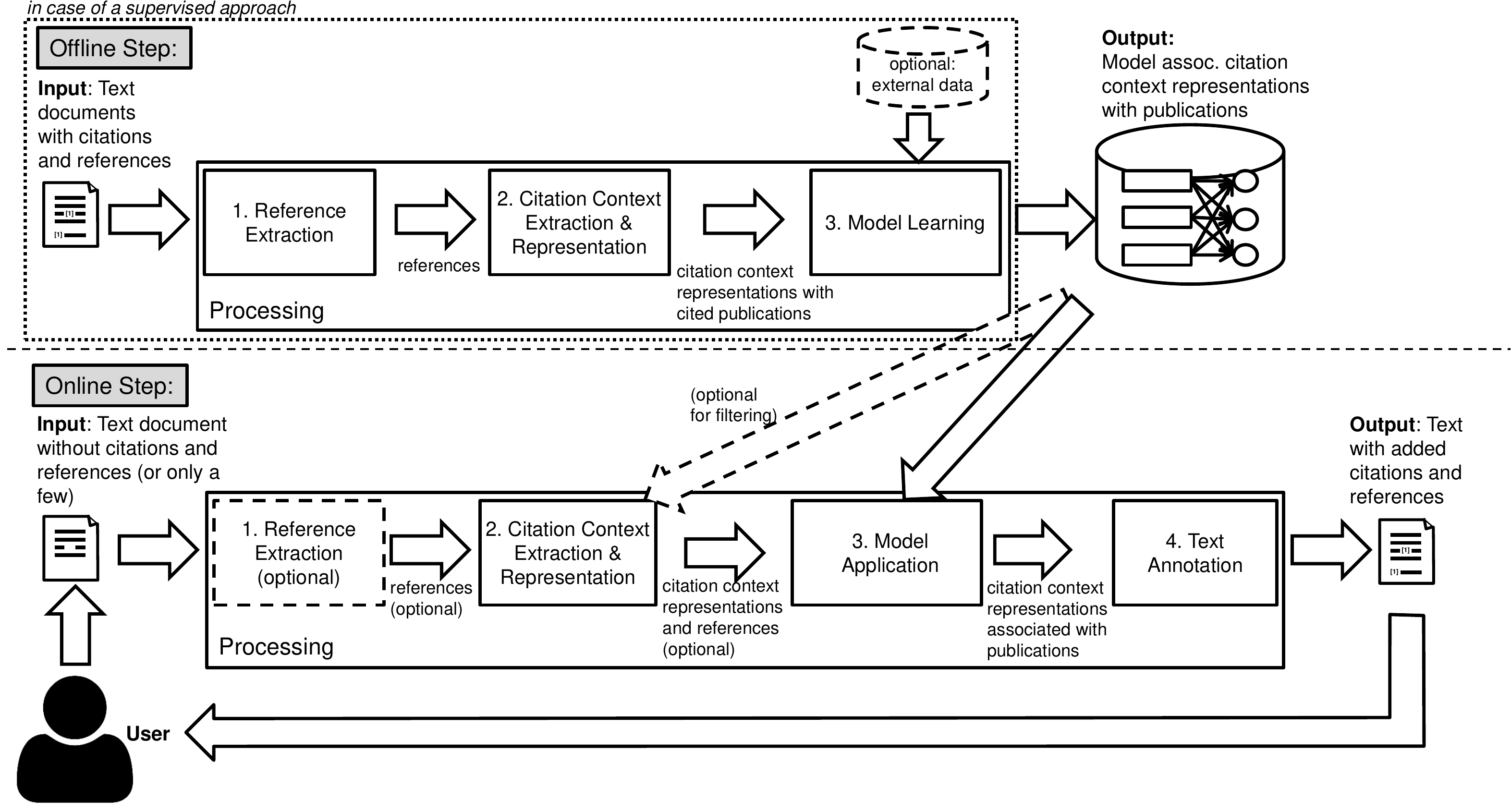}
 \caption{Architecture of a prototypical citation recommendation system.} 
 \label{fig:citation-recommendation-task}
\end{figure*}

In the following, we define local citation recommendation. By considering the whole document, abstract, or title as citation context, this definition can also serve as definition for global citation recommendation.
The general architecture of a context-aware citation recommendation system is depicted in Fig.~\ref{fig:citation-recommendation-task}. State-of-the-art citation recommendation approaches are supervised learning approaches. Thus, we can distinguish between an \textit{offline step} (or \textit{training phase} in machine learning setups), in which a recommendation model is learned based on a collection of documents, and an \textit{online step} (or \textit{testing/application phase}), in which the recommendation model is applied to a new incoming text document. Note, however, that unsupervised learning approaches and rule-based approaches are also possible (although, to date, to the best of our knowledge, none such have been proposed). In that case, the learning phase in the offline step is eliminated and a given model (e.g., set of rules) can be directly applied (see Fig.~\ref{fig:citation-recommendation-task}).

In the following, we give an overview of the steps in case of supervised learning (using the symbols summarized in Table~\ref{tab:symbols-task-definition}). 
Note that existing citation recommendation approaches use, to the best of our knowledge, content-based filtering techniques and are not based on other recommendation techniques, such as collaborative filtering or hybrid models. It is therefore not surprising that the approaches are mostly not personalized\footnote{Exceptions are \cite{Yin2017APWeb,Liu2013AIRS}, which also use the citing paper's author information besides the content.} (i.e., not incorporating user profiles).
Hence, our task formalization does not consider personalization. 

\begin{table}[tb]
\caption{Symbols used for formalizing \textit{citation recommendation}, grouped by the \textit{offline step} and the \textit{online step}.}
\label{tab:symbols-task-definition}
\begin{small}
\begin{tabular}{lp{4cm}}
\toprule
\textbf{Symbol}                      & \textbf{Description}                                      \\
\midrule
$D=\{d_1,...,d_i,...,d_n\}$      & set of citing documents in the offline step                                              \\
$R=\{r_1,...,r_m,...,r_M\}$      & references of all citing documents $D$                         \\
$C_i=\{c_{i1},...,c_{ij},...,c_{iN}\}$  & citation contexts from document $d_i$                     \\
$Z_i=\{z_{i1},...,z_{ij},...,z_{iN}\}$  & abstract citation contexts from document $d_i$                           \\
$Z$                           & set of all abstract citation contexts of $D$                            \\
$f$                           & mapping function                                                      \\
$g$                           & mapping function                                                      \\
\midrule
$d$                           & input document in the online step                                         \\
$R^d$                          & references of document $d$                                              \\
$C^d=\{c_{1}^d,...c_{k}^d, ..., c_{K}^d\}$ & potential citation contexts of document $d$                             \\
$Z^d=\{z_{1}^d,...,z_{k}^d,...,z_{K}^d\}$  & abstract representations of potential citation contexts of document $d$ \\
$R_{z_{k}^d}$                        & set of papers recommended for citation                                       \\
$d'$                          & input document $d$ enriched by recommended citations                        \\
\bottomrule
\end{tabular}
\end{small}
\end{table}

\subsubsection{Offline Step}
\paragraph{Input} Input is a set of documents $D = \{ d_1,...,d_n \}$, which we call in the following the \textit{citing documents}, with citations and references.\footnote{It should be noted that citation recommendation can be defined both on a citation context-level and on a document level. We here consider the task on a document level, because this enables us to have a more generic definition.}
\paragraph{Processing} The processing of the input texts consists of 
the following steps: 

\textbf{(1) Reference Extraction.}
All references from the reference sections of all citing documents are extracted and stored in a global index $R$. 

\textbf{(2) Citation Context Extraction \& Representation.}
First, all citation contexts $c_{ij} \in C_i$ from each citing document $d_{i}$ need to be extracted. 
Then, these citation contexts are transformed into the desired representation form (e.g., embedding vectors, bag-of-entities, etc.) $z_{ij}$: 
$$\forall d_{i} \in D ~ \forall c_{ij} \in C_{i}: c_{ij} \rightarrow z_{ij}$$

\textbf{(3) Model Learning.} 
Given the output of the previous steps (the citing documents $D$, the cited documents $R$, and the abstract citation contexts $Z$), we can learn a mapping function $f$ which maps each citation context representation 
$z_{ij}$ and its citing document $d_{i}$
to a reference (cited document) $r_{m} \in R$ 
as given by the training data:
$$
 \forall 
 z_{ij} \in Z ~ \forall d_{i} \in D ~~\quad~~ f: 
 (z_{ij}, d_{i})
 \rightarrow 
 r_{m} ~~~~ 
$$
Note that some approaches to citation recommendation might not use any other information from the citing documents besides the citation contexts, eliminating thus $d_{i}$ as argument in the mapping function. In those cases, only the representation of the citation context $z_{ij}$ is decisive (e.g., representation of a concept). 

The mapping function $f$ and the whole task can be formulated as a binary classification task (as also presented in \cite{Rokach2013}), especially in order to employ statistical models. 
Then, each citable document $r_{m}$ is considered as a class and the task is to determine if $(z_{ij}, d_i)$ should be in class $r_{m}$:
$$
 g(z_{ij}, d_{i}, r_{m}) \rightarrow [0,1]
$$
$[0, 1]$ is the probability of citing 
$r_{m}$ given $z_{ij}$ and $d_{i}$. As mentioned above, $d_{i}$ might be optional for some approaches. In reality, $g$ is often learned based on machine learning. However, one can also think of other ways to create $g$ (e.g., rule-based approaches). 

\paragraph{Output} Output is the function $g$, given the abstract citation contexts $Z$, the citing documents $D$, and the cited documents $R$.

\subsubsection{Online Step}

\paragraph{Input} Input is a text document $d$ without citations and references (or only a few ones).

\paragraph{Processing} Processing the document $d$ consists of the following steps: 

\textbf{(1) Reference Extraction (optional):} 
If $d$ already contains citations and a reference section, the \textit{references} $R^{d}$ from $d$ can be extracted and the corresponding representations can be retrieved from the database of cited papers $R$. These representations can be utilized for improving the citation recommendation within \textit{Model Application}, e.g., for a better topical coherence among existing and recommended citations \cite{Kobayashi2018}.  

\textbf{(2) Citation Context Extraction \& Representation:} 
First, if the existing \textit{citations} in document $d$ are to be used, the task is to extract and represent them in the same way as in the Offline Step.
Then, all \textit{potential citation contexts} $c_k^{d} \in C^{d}$ -- i.e., contexts in $d$, which are judged as suitable for having a citation -- are extracted from $d$ and transformed into the same abstract representation form $z_{k}^{d}$ as used in the \textit{Model Learning}:~
$\forall c_{k}^{d} \in C^{d} ~~:~~ c_{k}^{d} \rightarrow z_{k}^{d}$.
Note that, sometimes, an additional filtering step filters out all potential citation contexts which are not worth considering. 

\textbf{(3) Model Application:} 
Here, the mapping function $g$, learned during the training, is applied on the potential citation context representations $z_{k}^{d}$ of document $d$ for recommending citations: 
$$
 \forall z_{k}^{d} \in Z^{d} ~\quad~ R_{z_{k}^{d}} = \{ r_{m} ~|~ r_{m} \in R \land g(z_{k}^{d}, d, r_{m}) \geq \theta \}
$$
$R$ is thereby the global index of ``citable'' papers (gathered during the offline step).  $R_{z_{k}^{d}}$ is the set of recommended cited papers. These papers were classified as \textit{cited} with a likelihood of at least $\theta$. 

\textbf{(4) Text Enrichment} Given the document $d$ and the set of recommendations $R_{z_{k}^{d}}$ for each citation context representation $z_{k}^{d}$, the running text of document $d$ gets enriched by the recommended citations and the reference section of $d$ gets enriched by the corresponding references.

\paragraph{Output} Output is the annotated document $d'$.

\subsection{Related Research Fields}
\label{sec:related-research-fields}

\subsubsection{Non-Scholarly Citation Recommendation}

Also, outside academia, there is a demand for citing written knowledge. 
We can mention three kinds of documents, which often appear in such scenarios 
as citing documents: 
encyclopedic articles, news articles, and patents. 
Citation recommendation approaches developed for the scholarly field can in principle also be applied to such fields outside academia. Note, however, that each of the use cases might bring additional requirements and challenges. 
The scholarly domain is characterized by the use of a particular vocabulary, thus making it hard to apply models (e.g., embeddings) that were pre-trained on other domains (e.g., news). In contrast, documents in the non-scholarly field, such as news articles, often do not have a (dense) citation network. This might make it harder to build metadata-based representations of the documents and to evaluate the recommender systems, because no co-citation network can be used for the evaluation (see the fuzzy evaluation metrics in Sec.~\ref{sec:evaluation-methods-citation-recommendation}). In the following, we outline specifically developed approaches for non-scholarly citation recommendation.

\textbf{Encyclopedic articles as citing documents: }
The English Wikipedia is nowadays already very rich and quite complete in the number of articles included, but still lacks citations in the range of (at least) hundreds of thousands~\cite{JackLHK14}. 
This lack of citations diminishes the potential of Wikipedia to be a reliable source of information.
Since in Wikipedia mainly news articles are cited~\cite{FetahuMNA16}, several approaches have focused on developing methods for recommending news citations for Wikipedia \cite{Mishra14,Mishra2016ECIR,FetahuMNA16,FetahuMA15}.

\textbf{News articles as citing documents: }
Peng et al. \cite{Peng2016} approach the task of citation recommendation for news articles. They use a combination of existing implicit and explicit citation context representations as well as 200 preselected candidate articles instead of hundreds of thousands per citation context.

\textbf{Patents as citing documents: }
Authors of patents need to reference other patents in order to show the context in which the patent is embedded. Thus, approaches for patent citation recommendation have been proposed \cite{Mahdabi14}.

\begin{table*}
\caption{Overview of tools for extracting in-text citations (i.e., references' metadata and citations' positions in the text) from scientific publications, sorted by publication year.}
\label{tab:pdf-extraction-tools-overview}
\begin{footnotesize}
\begin{tabular}{P{3.7cm}P{2.7cm}P{1.2cm}P{1.9cm}P{2.8cm}P{2.3cm}}
\toprule 
Tool & Approach & Input format & Output format & Extracts citation contexts (citation context length) & Extracts citing paper's abstract  \\
\toprule 
CERMINE \cite{Tkaczyk2015}
& CRF   & pdf & xml & yes (300 words) & yes \\\midrule
ParsCit \cite{Councill2008}
& CRF   & txt & xml, txt & yes (200 words) & no \\ \midrule 
GROBID \cite{Lopez09Grobid,Lopez2015Grobid}
& CRF & pdf & xml  & no & yes  \\\midrule 
PDFX \cite{Constantin2013}
& rule-based   & pdf & xml & yes (300 words) & yes \\ \midrule 
Crossref pdf-extractor \cite{CrossRefpdfextract}
& rule-based & pdf & xml, bib   & no  & no  \\  \midrule 
IceCite \cite{Bast2017} 
& rule-based  & pdf & tsv, xml, json & no & yes  \\ 
\midrule 
Science Parse~\cite{Scienceparse} & CRF & pdf & json & yes & yes \\
\bottomrule
\end{tabular}
\end{footnotesize}
\end{table*}

\subsubsection{Scholarly Data Recommendation}

Scientists are not only confronted with an information overload regarding publications, but also regarding various other items, such as books, venues, and data sets. As a consequence, these items can also be recommended appropriately in order to assist the scientist in her work. 
Among others, approaches have been developed for recommending books \cite{Mooney2000}, scientific events \cite{Klamma2009}, venues \cite{Yang2012} and reviewers~\cite{Liu2014Rec} for given papers, patents \cite{Oh2013},
scientific data sets \cite{Singhal2013}, potentially identical texts (by that means identifying plagiarism) \cite{Gipp2014}, and newly published papers, via notifying functions \cite{Faensen2001}.

\subsubsection{Related Citation-based Tasks} 
\label{sec:related-citation-based-tasks}

In the following, we describe some citation-based tasks that are either strongly related to or an integral part of citation recommendation.

\paragraph{Citation network analysis} 
Citation network analysis describes the task of analyzing the references between documents in order to make statements about the scientific landscape and to investigate quantitatively scientific publishing. 
Among others, citation network analysis has been performed to determine communities of researchers~\cite{Yang2009KDD,Danon2005}, to find experts in a domain~\cite{Hashemi2013CIKM}, to know which researchers or publications have been or will become important, and to obtain trends in what is published over time~\cite{He2009CIKM}.
Note that citation network analysis operates on the document level and generally does not consider the document's contents. 

\paragraph{Citation context detection and extraction}
Each citation is textually embedded in a \textit{citation context}. 
The citation context can vary in length, ranging typically from a part of a sentence to many sentences. As shown in several analyses \cite{Alvarez2016,Abu-JbaraR12}, precisely determining the borders of the citation context is non-trivial. This is because several citations might appear in the same sentence and because citations can have different roles. While in some cases a claim made by the author needs to be backed up, in other cases a single concept (e.g., method, data set, or other domain-specific entity) needs to be referenced by a corresponding publication \cite{Faerber2019TPDL}.
In conclusion, there seems to be no consistent single optimal citation context length \cite{Alvarez2016,Ritchie2008,Ritchie2009}. 
Different citation context lengths have been used for citation recommendation (see Table~\ref{tab:comparison-approaches-part1}). 

To extract citation contexts and references from papers, specific approaches have been developed \cite{DBLP:journals/corr/abs-1802-01168,Tkaczyk2018}. 
These approaches were developed for PDFs with a paper-typical layout. 
They are not only capable of extracting a paper's metadata, such as title, author information, and abstract, in a structured format, but also the references from the reference section, as well as linking the citation markers in the text to the corresponding references.
Table~\ref{tab:pdf-extraction-tools-overview} provides an overview of the  existing publicly available implementations for extracting in-text citations from scientific papers. 
Note that we limited ourselves to implementations which were designed for scientific papers as input and which are still deployable; other PDF extraction tools are not considered by us (see \cite{Bast2017,DBLP:journals/corr/abs-1802-01168,Tkaczyk2018} for an overview of further PDF-to-text tools). Furthermore, we excluded tools, such as Neural ParsCit~\cite{Prasad2018NeuralParscit}, which do not output the positions of the citations in the text.
Given these tools, we can observe the following: 
(1)~All underlying approaches are a rule engine or a conditional random field. 
(2)~Several tools (e.g., ParsCit) have the additional feature that they can extract not only the fulltext from the PDF documents, but also a citation context around the found citation markers.
(3)~Several tools (e.g., ParsCit) require plaintext files as input. Transforming PDF to plaintext is, however, an additional burden and leads to noise in the data.
(4)~The tools differ considerably in the processing time needed for processing PDF files \cite{Bast2017}. ParsCit and GROBID, which have been used most frequently by researchers, to our knowledge, are among the fastest.

\paragraph{Citation context characterization}
Citations can have different roles, i.e., citations are used for varying purposes. These reasons are also called \textit{citation functions}. The citation function can be determined -- to some degree automatically -- by analyzing the citation context and by extracting features \cite{Teufel2009annotation,Moravcsik1975,Teufel2006EMNLP}. Similar tasks to the citation function determination are the polarity determination (i.e., if the author speaks in a positive, neutral, or negative way about the cited paper) \cite{Abu-Jbara2013,GhoshD017} and the determination of the citation importance~\cite{Valenzuela2015,Chakraborty2016}.

The general typical structure of publications has been studied and brought into a schema, such as the IMRaD structure~\cite{Sollaci2004}, standing for introduction, methods, results, and discussion. In \cite{Bertin2016}, for instance, the authors find out that the average number of citations among the same sections in article texts is invariant across all considered journals, with the introduction and discussion accounting for most of the citations. Furthermore, apparently the age of cited papers varies by section, with references to older papers being found in the methods section and citations to more recent papers in the discussion. 
Although such insights have not been used for development of citation recommendation approaches yet, we believe that they can be beneficial for better approximating real human citing behavior. 

\paragraph{Citation-based document summarization} 
Citation-based document summarization is based on the idea that the citation contexts within the citing papers are written very carefully by the authors and that they reveal noteworthy aspects of the cited papers. Thus, by collecting all citation contexts and grouping them by cited papers, summaries and opinions about the cited papers can be obtained, opening the door for citation-based automatic survey generation and automatic related work section generation \cite{Abu-Jbara2011,Elkiss2008,Mohammad2009}. 

\paragraph{Citation matching and modeling}
Citation matching~\cite{Pasula2002} deals with the research challenge of finding identical citations in different documents in order to build a coherent citation network, i.e., a global index of citations for a document collection. 

Representing the metadata of both citing and cited papers in a structured way is essential for any citation-based task. 
Recently, several ontologies, such as FaBiO and CiTO \cite{Peroni2012}, have been proposed for this purpose. 
Besides the metadata of papers, further relations and concepts can be modeled ontologically in order to facilitate transparency and advances in research~\cite{Pertsas2017}.

\section{Comparison of Citation Recommendation Approaches}
\label{sec:comparison-approaches}

Approaches to (local and global) citation recommendation have been published over the years, using diverse methods, and proposing many variations of the citation recommendation task, such as a recommendation across languages \cite{Tang2014SIGIR} or using specific metadata about the input text \cite{Rokach2013,Ebensu2017}. However, no overview and comparison of these approaches has been presented in the literature so far. In the following, we give such an overview.

\subsection{Corpus Creation} 
\label{sec:methodology-collection-dimensions}

\begin{table}
\centering
\caption{Approaches to global and local citation recommendation (CR).}
\label{tab:all-approaches-citerec}
\begin{footnotesize}
\begin{tabular}{ll@{}c}
\toprule
Reference & Venue & Local CR \\ 
\midrule
McNee et al. \cite{McNee2002} & {CSCW}'02 &  \\
Strohman et al. \cite{Strohman2007SIGIR} & {SIGIR}'07 &  \\
Nallapati et al. \cite{Nallapati2008} & KDD'08 &  \\ 
Tang et al. \cite{Tang2009PAKDD} & {PAKDD}'09 &  \\
He et al. \cite{He2010WWW} & {WWW}'10 & \checkmark \\
Kataria et al. \cite{Kataria2010} & AAAI'10 & \checkmark \\ 
Bethard et al. \cite{Bethard2010} & CIKM'10 &  \\ 
He et al. \cite{He2011WSDM} & WSDM'11 & \checkmark \\
Lu et al. \cite{Lu2011CIKM} & CIKM'11 &  \\
Wu et al. \cite{Wu2012FSKDn} & FSKD'12 &  \\
He et al. \cite{He2012SPIRE} & SPIRE'12 & \checkmark \\
Huang et al. \cite{Huang2012CIKM} & CIKM'12 & \checkmark \\
Rokach et al. \cite{Rokach2013} & LSDS-IR'13 & \checkmark \\
Liu et al. \cite{Liu2013AIRS} & AIRS'13 & \checkmark \\
Jiang et al. \cite{Jiang2013} & TCDL Bulletin'13 &  \\
Zarrinkalam et al. \cite{Zarrinkalam2013Program} & Program'13 &   \\
Duma et al. \cite{Duma2014ACL} & ACL'14 & \checkmark \\
Livne et al. \cite{Livne2014SIGIR} & SIGIR'14 & \checkmark \\
Tang et al. \cite{Tang2014SIGIR} & SIGIR'14 & \checkmark \\
Ren et al. \cite{Ren2014KDD} & KDD'14 &  \\
Liu et al. \cite{Liu2014JCDL} & JCDL'14 & \\
Liu et al. \cite{Liu2014CIKM} & CIKM'14 &  \\
Jiang et al. \cite{Jiang2014WebKR} & Web-KR'14 &  \\
Huang et al. \cite{HuangWCMG15} & WCMG'15 & \checkmark \\
Chakraborty et al. \cite{Chakraborty2015ICDE} & ICDE'15 &  \\
Hsiao et al. \cite{Hsiao2015MDM} & MDM'15 &  \\
Gao et al. \cite{Gao2015FSKD} & FSKD'15 &  \\
Lu et al. \cite{Lu2015APWeb} & APWeb'15 &  \\
Jiang et al. \cite{Jiang2015CIKM} & CIKM'15 &  \\
Liu et al. \cite{Liu2016iConf} & iConf'16 &  \\
Duma et al. \cite{Duma2016LREC} & LREC'16 &  \\
Duma et al. \cite{Duma2016DLib} & D-Lib'16 &  \\
Yin et al. \cite{Yin2017APWeb} & APWeb'17 & \checkmark \\
Ebesu et al. \cite{Ebensu2017} & SIGIR'17 & \checkmark \\
Guo et al. \cite{GuoCHMFY17} & IEEE'17 &  \\
Cai et al. \cite{Cai2018AAAI} & AAAI'18 &  \\
Bhagavatula et al. \cite{Bhagavatula2018} & NAACL'18 & \\
Kobayashi et al. \cite{Kobayashi2018} & JCDL'18 & \checkmark \\
Jiang et al. \cite{Jiang2018JCDL} & JCDL'18 &  \\
Han et al. \cite{Han2018ACL} & ACL'18 & \checkmark \\
Jiang et al. \cite{Jiang2018SIGIR} & SIGIR'18 &  \\
Zhang et al. \cite{Zhang2018ISMIS} & ISMIS'18 &  \\
Cai et al. \cite{Cai2018IEEE} & IEEE TLLNS'18 &  \\
Yang et al. \cite{Yang2018JIFS} & JIFS'18 &  \\
Dai et al. \cite{Dai2018JAIHC} & JAIHC'18 &  \\
Yang et al. \cite{Yang2018IEEEAccess} & IEEE Access'18 & \checkmark \\
Mu et al. \cite{Mu2018IEEE} & IEEE Access'18 &  \\
Jeong et al. \cite{Jeong2019arxiv} & arXiv'19 & \checkmark \\
Yang et al. \cite{Yang2019IEEE} & IEEE Access'19 &  \\
Dai et al. \cite{Dai2019IEEEAccess} & IEEE Access'19 &  \\
Cai et al. \cite{Cai2019IEEE} & IEEE Access'19 &  \\
\bottomrule
\end{tabular}
\end{footnotesize}
\end{table}

Following a similar procedure as in~\cite{Beel2016}, we collect the papers for our comparison as follows:
\begin{enumerate}
 \item On May 3, 2019, we searched in DBLP for papers containing ``citation'' and ``rec*'' in the title. This resulted in a set of 179 papers. We read those papers and manually classified each of them whether they present an approach to (local or global) citation recommendation or not. 
 \item In a further step, we also investigated all papers referenced by the so-far given relevant papers,  and the ones that refer to these so-far given papers, and classify them as relevant or not. 
 \item To avoid missing any papers, we used Google Scholar as an academic search engine with the query keywords ``citation recommendation'' and ``cite recommend,'' as well as the Google Scholar profiles from the authors of the so-far relevant papers. Based on that, we added a few more relevant papers to our corpus.\footnote{\cite{Rokach2013,Liu2016iConf} are papers which are not indexed in DBLP, but which can be found on Google Scholar or Semantic Scholar.}
\end{enumerate}
Overall, 51 papers propose a novel, either global or local citation recommendation approach (see Table~\ref{tab:all-approaches-citerec}).
Out of these, 17 present \textit{local} citation recommendation approaches, that is, approaches that use a specific citation context within the input document (see Sec.~\ref{sec:terminology} for the distinction between local and global citation recommendation). 
This means that only 33.3\% of the approaches denoted by the corresponding authors as \textit{citation recommendation approaches} are actually designed for using citation contexts as input and are therefore truly citation recommendation approaches (see Sec.~\ref{sec:terminology}).

Note that we consider only papers presenting approaches to citation recommendation, and not those on data analysis (e.g., citation graph analysis). 
We also do not consider papers presenting approaches for recommending papers 
that do not use any text as the basis for the recommendation, but instead use other information, such as the papers' metadata.

\subsection{Corpus Characteristics}

Table~\ref{tab:all-approaches-citerec} lists all 51 papers on citation recommendation, together with the papers' venues and an indication of whether the described approach targets local or global citation recommendation.
We can point out the following findings regarding the evolution of these approaches over time: 

\begin{figure}
    \centering
    \includegraphics[width=\linewidth]{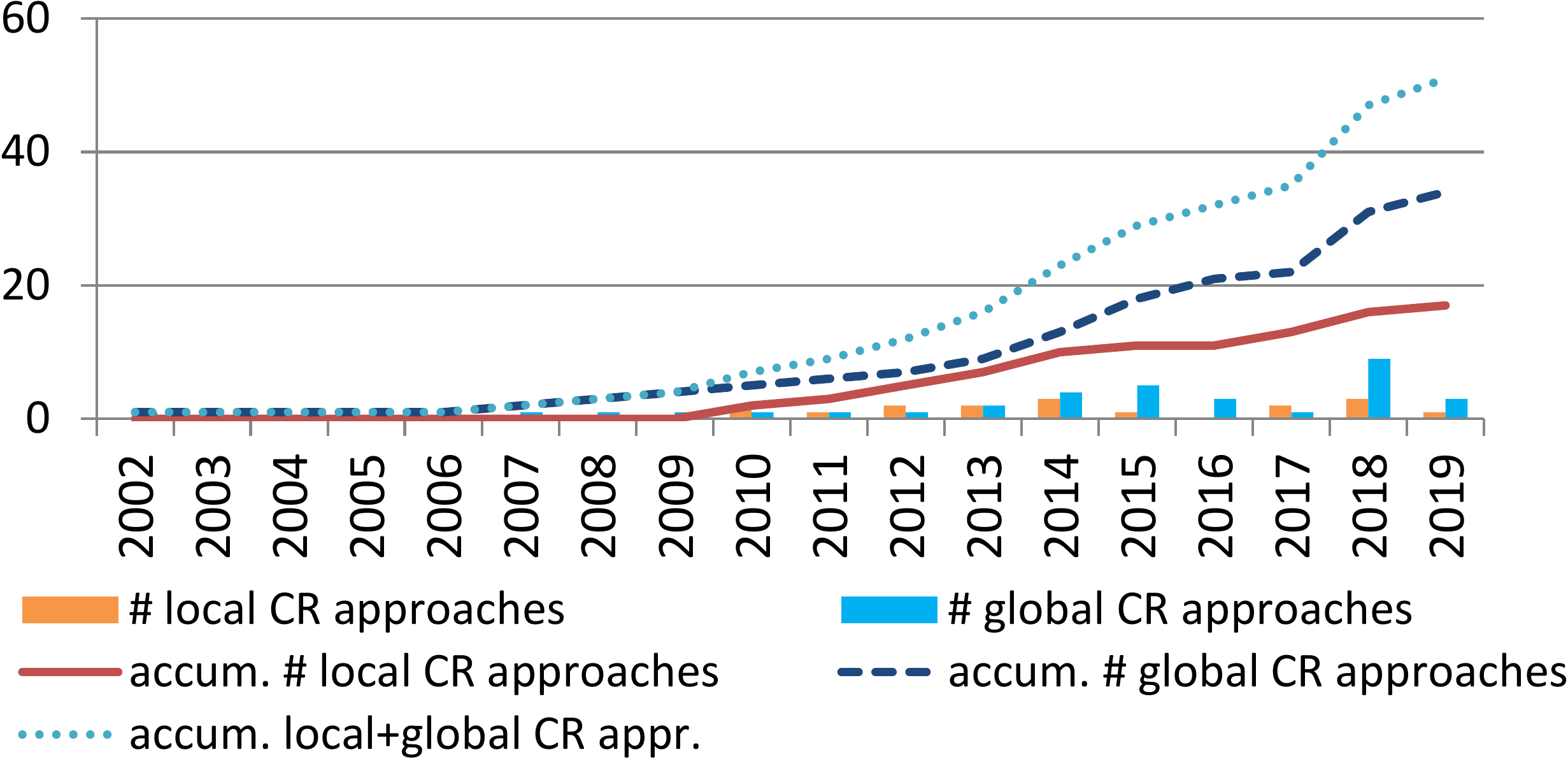}
    \caption{Frequencies of citation recommendation (CR) approaches by publication year.}
    \label{fig:approaches-freq} 
\end{figure}

\begin{figure*}[tb]
 \centering
 \includegraphics[width=0.9\textwidth]{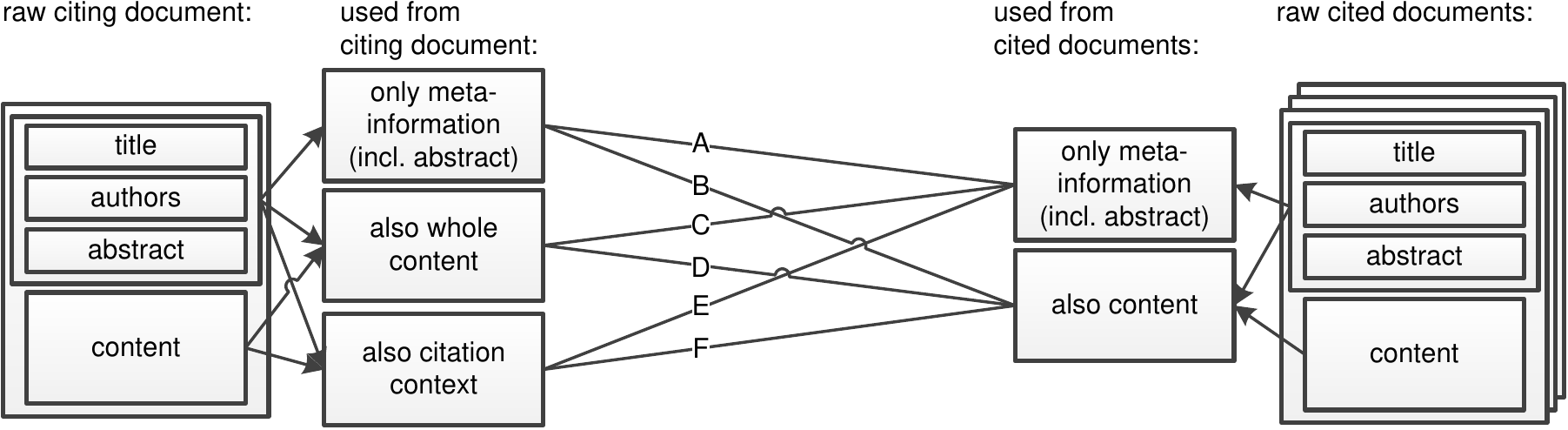}
 \caption{Classification of citation recommendation approaches based on their set-up. The approaches are classified as follows: A: \cite{McNee2002,Ren2014KDD,Jiang2014WebKR,Jiang2015CIKM,Hsiao2015MDM,Chakraborty2015ICDE,Gao2015FSKD,Lu2015APWeb,GuoCHMFY17,Bhagavatula2018,Jiang2018JCDL,Jiang2018SIGIR,Cai2018AAAI,Cai2018IEEE,Yang2018JIFS,Dai2018JAIHC,Mu2018IEEE,Yang2019IEEE,Cai2019IEEE}: B: \cite{Bethard2010}; C: \cite{Zarrinkalam2013Program}; D: \cite{Strohman2007SIGIR,Nallapati2008,Tang2009PAKDD,Wu2012FSKDn,Jiang2013,Liu2014CIKM,Liu2016iConf,Duma2016LREC,Duma2016DLib,Zhang2018ISMIS,Dai2019IEEEAccess,Lu2011CIKM,Liu2014JCDL}; E: \cite{He2010WWW,He2011WSDM,Huang2012CIKM,Rokach2013,Liu2013AIRS,Duma2014ACL,Tang2014SIGIR,HuangWCMG15,Yin2017APWeb,Ebensu2017,Yang2018IEEEAccess,Jeong2019arxiv}; F: \cite{He2012SPIRE,Livne2014SIGIR,Kobayashi2018,Han2018ACL,Kataria2010}
 The numbers correspond to the references in the reference section.}
 \label{fig:related-work-overview-diagram}
\end{figure*}

\begin{enumerate}
 \item We can observe that approaches to citation recommendation have been published over the last 17 years (see Fig.~\ref{fig:approaches-freq}). The task of \textit{global} citation recommendation has attracted the interest of researchers at an earlier stage than local citation recommendation (first publication year 2002 \cite{McNee2002} vs. 2010 \cite{He2010WWW}). Both the number of approaches to global citation recommendation and local citation recommendation has increased continuously. Overall, more approaches to global citation recommendation system have been published than approaches to local citation recommendation. However, note that the most recent publications on global citation recommendation have been published in very short time intervals at similar or same venues from partially identical authors (see Table~\ref{tab:all-approaches-citerec}). 
 \item Some precursor works on the general task of analyzing and predicting links between documents \cite{Cohn2000} have been published since 2000, while
 global citation recommendation has been targeted by researchers since 2002. Among others, there might be two major aspects that can explain the emergence of citation recommendation approaches at that time. Firstly, the number of papers published per year has increased exponentially. It became common in the 2000s to publish and to read publications online on the Web. Secondly, citations have become disproportionately more common over the years, that is, the number of citations has increased faster than the number of publications. Comparing the five-year periods 1999/2003  and 2004/2008 in \cite{NSF2014}, the number of publications increased by 33\%, while citations increased by 55\%.
 \item Before the content-based (local and global) citation recommendation approaches -- as considered in this survey --, several systems had already been proposed that use purely the citation graph as basis for the recommendation. This ``prehistory'' of content-based citation recommendation is explainable by the fact that quantitative science studies such as bibliometrics have a long history, and were already quite established in the 2000s. 
 \item Having an appropriate and large collection of scientific papers as evaluation and training data is crucial and not easy to obtain, since -- especially in the past -- papers were often ``hidden'' behind paywalls of publishers. Therefore, it is not very surprising that several approaches \cite{Bethard2010,Jiang2015CIKM,Bhagavatula2018,Jiang2018JCDL} consider only abstracts as citing documents instead of the papers' content. Citation recommendation then turns into reference recommendation for abstract texts. 
 \item Citation recommendation is located in the intersection of the research areas \textit{information retrieval}, \textit{digital libraries}, \textit{natural language processing}, and \textit{machine learning}. This is also reflected in the venues in which approaches to citation recommendation have been presented. Considering both global and local citation recommendation, SIGIR, IEEE Access, CIKM, and JCDL have been chosen most frequently as venues (5 times SIGIR, 5 times IEEE Access, 5 times CIKM, 3 times JCDL; together accounting for 35\% of all papers). Particularly, IEEE Access has become popular as a venue for publishing citation recommendation approaches by a few researches in 2018 and 2019. Note that this journal's reviewing and publication process is designed to be very tight (one review round takes 7 days) and that IEEE has an article processing charge. Our paper corpus also contains a few publications from medium-ranked conferences, such as AIRS \cite{Liu2013AIRS}. It became apparent that these papers provide less comprehensive evaluations, but relatively high evaluation results (see the  \textit{evaluation metrics} paragraph in Section \ref{sec:approaches-comparison-sub}). Due to missing baselines, these results need to be taken with care.
 \item Considering purely local citation recommendation, SIGIR (3 times) and ACL (2 times) occur most frequently as venue. The remaining venues occur only once.
\end{enumerate}

\begin{figure}
    \centering
    \includegraphics[angle=90,origin=c,width=0.63\linewidth]{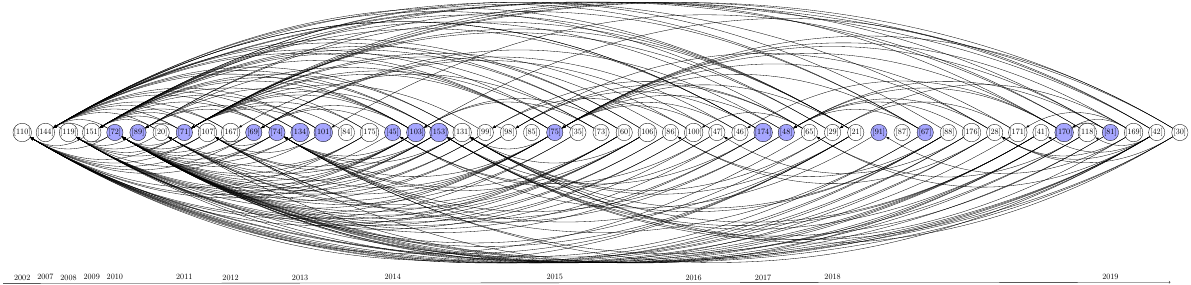} 
    \caption{Citations between papers presenting citation recommendation approaches. Local citation recommendation approaches are highlighted in blue.}
    \label{fig:fig-cit}
\end{figure}

\textbf{Big picture.}
In Fig.~\ref{fig:related-work-overview-diagram}, we present visually a ``big picture'' of the different settings in all citation recommendation approaches. 
We thereby differentiate between what data is used from the \textit{citing} documents (either only metadata (incl. abstract), or metadata plus content, or metadata plus specific citation contexts), and what data is used from the \textit{cited} documents (either only metadata, or metadata plus content). 
Note that approaches using the metadata or the content of the citing documents make up the group of global citation recommendation approaches, while approaches using specific citation contexts target local citation recommendation.
Note also that approaches using only the metadata of the citing documents can be regarded as targeting both the expert setting and the non-expert setting (see Section~\ref{sec:scenarios-citation-recommendation}), while the other approaches are designed primarily for the expert setting.
The publications that propose the approaches sometimes do not point out in detail what data is used (e.g., whether the author information of the citing papers is also used), which makes a valid comparison infeasible. 
Thus, this ``big picture'' figure tries to provide a clear picture of what has been pursued so far. 
Notable, for instance, is that 23.5\% (12 out of 51) of all approaches use citation contexts (less than the whole content) of the citing documents and only the metadata of the cited documents (see class E). In contrast, we can find only one approach that uses the whole content of the citing documents and only the metadata of the cited documents (see class C).
We can mention two potential reasons for this fact. Firstly, it can be difficult to obtain the publications' full texts (due to, among other reasons, limited APIs and copyright issues). Secondly, operating only with papers' metadata is also easier from a technical perspective.

\textbf{Citation relationships.} 
Fig.~\ref{fig:fig-cit} shows the citation-relationships between papers with citation recommendation approaches. 
The papers are thereby ordered from left to right by publishing year. 
It is eye-catching that there is no continuous citing behavior along the temporal dimension, i.e., a paper in our set does not necessarily cite preceeding papers in our set. However, in some cases we can explain this by the fact that publications were published within short time intervals. Consequently, the authors might not have been aware of other approaches which had either been published very recently or had not yet been published.
Nevertheless, we can observe that authors of citation recommendation approaches do omit references to other citation recommendation approaches.

\subsection{Comparison of Local Citation Recommendation Approaches}
\label{sec:approaches-comparison-sub}

When comparing citation recommendation approaches, it is important to differentiate between approaches to local citation recommendation (making recommendations based on a small text fragment) and approaches to global citation recommendation. To understand that, consider a scenario in which a text document with 20 citation markers is given. In case of local citation recommendation, it is not uncommon to provide, for instance, three recommendations per citation context. 
However, a global citation recommendation system would provide only a list of 60 recommendations without indications where to insert the corresponding citation markers. 
In our mind, it is not reasonable to call this process context-aware citation recommendation and to evaluate the list of 60 recommendations in the same way as the 20 lists with 3 recommendations, since citations are meant to back up single statements and concepts on a clause level, i.e., being suitable only for specific contexts. 
Note also that global recommendation approaches in the context of paper recommendation are covered by existing surveys (see Introduction). This survey, in contrast, focuses on context-awareness, which, to date, has not yet been considered systematically.
Thus, in this subsection, we compare only the 17 approaches to local citation recommendation. 

In order to characterize and distinguish the different approaches from each other, we introduce the following dimensions: 
\begin{enumerate}
 \item What is the underlying approach and to which data mining technique is it associated?
 \item What information is used for the user modeling, if any?
 \item Is the set of candidate papers prefiltered before the recommendation?
 \item What is used as the citation context (e.g., 1 sentence or 50 words before and after the citation marker)? 
 \item Is the citation context pre-specified in the evaluation or do cite-worthy contexts first need to be determined by the algorithm? 
 \item Is the content of the cited papers also needed (limiting the evaluation to corresponding data sets)?
 \item Which evaluation data set is used (e.g., CiteSeerX or own data set)?
 \item From which domain are the papers used in the evaluation (e.g., computer science)?
 \item What are the used evaluation metrics?
\end{enumerate}
Table~\ref{tab:comparison-approaches-part1} shows the classification of the approaches according to these dimensions. 
While in the following we point out the main findings per dimension, note that we also provide a description of the single approaches and their characteristics in an online semantic  wiki.\footnote{\url{http://wiki.citation-recommendation.org}.} 

\begin{sidewaystable*}
\vspace*{17.1cm}
\hspace*{-2cm}
\centering
\caption{Overview of local scientific citation recommendation approaches, listed chronologically by publication date (considering year and month).} 
\label{tab:comparison-approaches-part1}
\begin{footnotesize}
\begin{tabular}{P{0.8cm}@{}P{0.8cm}@{}P{0.95cm}@{}P{3.25cm}P{1.05cm}P{0.5cm}P{2.5cm}P{1.15cm}P{1.9cm}P{2.5cm}P{1.6cm}P{2.95cm}}
\toprule
Pa\-per & Year & Group & Approach & User model & Pre\-filter & Citation context length & Citation placeholders & Cited papers' content needed & Evaluation data set & Domain & Evaluation metrics \\
\toprule
\cite{He2010WWW} & 2010 & b &  Probabilistic model (Gleason's Theorem) & -- & -- & 50 words before and after & yes & no & CiteSeerX & Computer science & recall, co-cited prob., nDCG, runtime \\
\midrule
\cite{Kataria2010} & 2010 & b & Topic model (adapt. LDA) & -- & -- & 30 words before and after & yes & yes & CiteSeer & Computer science & RKL \\
\midrule
\cite{He2011WSDM} & 2011 & a & Ensemble of decision trees & -- & -- & 50 words before and after & no & no & CiteSeerX & Computer science & recall, co-cited probability, nDCG \\
\midrule
\cite{He2012SPIRE} & 2012 & c & Machine translation & -- & -- & 1 sentence & yes & yes & Own dataset & Computer science & MAP \\
\midrule
\cite{Huang2012CIKM} & 2012 & c & Machine translation & -- & -- & 1-3 sentences & yes & no & CiteSeer \& CiteULike & Computer science & precision, recall, F1; Bpref, MRR \\
\midrule
\cite{Rokach2013} & 2013 & a & Ensemble of supervised ML techniques & author & top 500 & 50 words before and after & yes & no & CiteSeer \& CiteULike & Computer science & F1, precision, runtime \\
\midrule
\cite{Liu2013AIRS} & 2013 & a & SVM & author & -- & on average 13.4 words & yes & no & Own dataset & Computer science & recall, MAP \\
\midrule
\cite{Duma2014ACL} & 2014 & a & cos similarity of vectors (TF-IDF based) & -- & -- & 5-30 words before and after & yes & depending on variant & Part of ACL Anthology & Comput. linguistics & accuracy \\
\midrule 
\cite{Livne2014SIGIR} & 2014 & a & Regression trees (gradient boosted) & author & top 500 & 50 words before and after & no & yes & Own dataset & Computer science & nDCG \\
\midrule
\cite{Tang2014SIGIR} & 2014 & d & Learning-to-rank & -- & -- & sentence plus sentence before and after & yes & no & Own dataset & Computer science and technology & recall, MAP, MRR \\
\midrule
\cite{HuangWCMG15} & 2015 & d & Neural network (feed-forward) & -- & var\-iable & sentence plus sentence before and after & yes & no & CiteSeer & Computer science & MAP, MRR, nDCG \\
\midrule
\cite{Yin2017APWeb} & 2017 & d & Neural network (CNN) & -- & var\-iable & sentence plus sentence before and after & yes & no (but title + abstract) & own (same as in \cite{Liu2013AIRS}) & Computer science & MAP, recall \\ 
\midrule
\cite{Ebensu2017} & 2017 & d & Neural network (CNN + RNN) & author & top 2048 & 50 words before and after & yes & no & RefSeer & Computer science & Recall, nDCG, MAP, MRR \\ 
\midrule
\cite{Kobayashi2018} & 2018  & d & cos. similarity of paper embeddings & -- & -- & 1 sentence & yes & yes & Own dataset (from ACM library) & Computer science & nDCG \\
\midrule
\cite{Han2018ACL} & 2018 & d & Dot product of 2 paper embeddings & -- & -- & 50 words before and after & yes & yes & NIPS, ACL-ANT, CiteSeer + DBLP & Computer science & recall, MAP, MRR, nDCG \\
\midrule
\cite{Yang2018IEEEAccess} & 2018 & d & Neural network (LSTM) & author, venue & -- & 5 sentences before and after & yes & yes & AAN + DBLP & Computer science & recall, MAP, MRR \\
\midrule
\cite{Jeong2019arxiv} & 2019 & d & Neural network (feed-forward) & -- & -- & 50 words before and after & yes & no & AAN + own dataset & Computer science & MAP, MRR, recall \\
\bottomrule
\end{tabular}
\end{footnotesize}
\end{sidewaystable*}

\begin{enumerate}
 \item \textbf{Approach: }
 A variety of methods have been developed for local citation recommendation. 
 We can group them into the following four groups: 
 
 \begin{enumerate}
  \item \textbf{Hand-crafted feature based models \cite{He2011WSDM,Rokach2013,Liu2013AIRS,Duma2014ACL,Livne2014SIGIR}.} All approaches in this group are based on features that were hand-crafted by the developers. Text similarity scores obtained between the citation context and the candidate papers are examples of text-based features. Remarkably, all features used for the approaches are kept comparably simple. Moreover, the approaches do not use additional external data sources, but rather statistics derived from the paper collection itself (e.g., citation count and text similarity). 
  Relatively basic techniques used for the ranking of citations for the purpose of citation recommendation (e.g., logistic regression and linear SVM \cite{Liu2013AIRS}, or merely the cosine similarity of TF-IDF vectors \cite{Duma2014ACL}) seem to lead to already noteworthy evaluation results and, thus, can serve as strong baselines for the evaluations of other systems.
  Among the most complex presented methods are an ensemble of decision trees \cite{He2011WSDM} and gradient boosted regression trees \cite{Livne2014SIGIR}. 
  Note, however, that their superiority compared to simpler models is hard to judge due to differing evaluation settings, such as data sets and metrics. 
  
  In recent years, no novel approaches of this group have been published any more (latest one from 2014), likely due to the fact that (1) the obvious features have already been used and evaluated, and (2) recent approaches (e.g., neural networks) seem to outperform the hand-crafted feature-based models. 
  Nevertheless, hand-crafted feature based models provide the following advantages: 
  \textit{1.~Scalability:} Since both the computation of the features and the used classifier/regression model are kept rather simple, the citation recommendation approaches become very scalable and fast.
  \textit{2.~Explainability:} The described techniques are particularly beneficial when it comes to getting to know which features are most indicative for recommending appropriate citations. 
  \textit{3.~Small data:} The models do not require huge data sets for training, but may already work well for small data sets (e.g., a few thousand documents).
  Existing approaches in this group use mainly lexical features and other bibliometrics-based features (e.g., citation count). Hand-crafted features focusing on the semantics and pragmatics of the citation contexts and of the candidate cited documents, are missing. In the future, one can envision a scenario in which claims or argumentation structures are extracted from the citation contexts and compared with the claims/argumentation structures from the citable documents. 
  
  \item \textbf{Topic modeling \cite{He2010WWW,Kataria2010}.} Topic modeling is a way of representing text (here: candidate papers and citation contexts) by means of abstract topics, and thereby exploiting the latent semantic structure of texts.
  Topic modeling became popular, among others, after the publication of the LDA approach by Blei et al. in 2002 and was applied to local citation recommendation in 2010 \cite{He2010WWW,Kataria2010}.
  Using topic modeling in the context of citation recommendation means to adapt default topic modeling approaches, which work purely on plain text documents, in such a way that they can deal with both texts and citations. 
  To this end, He et al.~\cite{He2010WWW} use a probabilistic model based on Gleason's Theorem, while Kataria et al.~\cite{Kataria2010} propose the LDA-variations Link-LDA and Link-PLSA-LDA. 
  
  Note that topic modeling per se is computationally rather expensive and may require more resources than approaches of the group (a). Moreover, conceptually it might be designed rather for longer texts, and, thus, more suitable for global citation recommendation (where it has been applied in \cite{Tang2009PAKDD,Nallapati2008}). 
  In the series of citation recommendation approaches, topic modeling has been applied within a relatively short time interval (2010 only for local citation recommendation; 2008 and 2009 in case of global citation recommendation) and has been replaced first by machine translation models (group (c)) and later by neural network-based approaches (group (d)). 
  
  \item \textbf{Machine translation \cite{He2012SPIRE,Huang2012CIKM}.} The authors of \cite{He2012SPIRE,Huang2012CIKM} apply the concept of machine translation to local citation recommendation.
  These approaches had been published also within a short time frame, namely only in 2012. 
  Using machine translation might appear surprising at first. However, the developed approaches do not translate words from one language into another, but merely ``translate'' the citation context into the cited document (written in the same language, but maybe with a partially different vocabulary). In this way, the vocabulary mismatch problem is avoided.
  The first published approach using machine translation for citation recommendation was designed for global citation recommendation \cite{Lu2011CIKM}. Here, the words in the citing document are translated to the words in the cited document. This requires the cited documents' content to be available. 
  Approaches to local citation recommendation follow: In \cite{He2012SPIRE}, the translation model uses several positions in the citable document for translations. However, this makes the approach computationally very expensive. 
  The last approach in this group \cite{Huang2012CIKM} translates the citing document merely into the identifiers of the cited documents and does not use the cited documents' content any more. By doing that, the authors obtain surprisingly high evaluation results. 
  Note that machine translation is a statistical approach and requires a large training data set. However, in the published papers and their evaluations, rather small data sets (e.g., 3,000 and 14,000 documents in \cite{Huang2012CIKM} and 30,000 documents in \cite{He2012SPIRE}) are used. 
  Moreover, high thresholds for the translation probability may be set to make the machine translation approach feasible \cite{Huang2012CIKM}. 
  
  \item \textbf{Neural networks \cite{Tang2014SIGIR,HuangWCMG15,Yin2017APWeb,Ebensu2017,Kobayashi2018,Han2018ACL}.} This group contains not only many approaches to local citation recommendation (6 out of 17, that is, 35\%), but also the most recent ones: here, papers have been published since 2014. 
  Due to the large field of neural network research in general, the architectures proposed here also vary considerably. 
  Although there are also relatively generic neural network architectures  applied~\cite{Yin2017APWeb,Tang2014SIGIR}, we can observe a tendency in increasing complexity of the approaches. Approaches are either specifically designed for texts with citations (e.g., \cite{Ebensu2017,Kobayashi2018}) or consider texts with citations as a special case of hyperlinked documents \cite{Han2018ACL}. In the first subgroup are approaches using convolutional neural networks \cite{Ebensu2017} and special attention mechanisms, such as for authors \cite{Ebensu2017}. In the latter subgroup is an approach which uses two vector representations for each paper. 
  Note that the approaches in this approach group do not incorporate any user model information, but work purely on the sequence of words. An exception is \cite{Ebensu2017} which exploits the citing document's author information. 
  
  When it comes to deciding whether neural networks should be used in a productive system, one should note that neural networks need to be trained on large data sets. 
  In recent years, large paper collections have been published (see Sec.~\ref{sec:data-sets}). However, also the infrastructure, such as GPUs, needs to be available. Moreover, considerable approximations need to be applied to keep the approach feasible. This includes the negative sampling strategy \cite{Yin2017APWeb,HuangWCMG15,Kobayashi2018,Han2018ACL}. 
  But also a pre-filtering step before the actual citation recommendation approach is often performed, which reduces the set of candidate papers significantly \cite{HuangWCMG15}. 
  
  Han et al.~\cite{Han2018ACL}, who propose one of the most recent citation recommendation systems and who evaluate their approach on data sets with real-world sizes, report recall@10 values of 0.16/ 0.32/ 0.21 and nDCG@10 values of 0.08/0.21/0.13 for the data sets NIPS, ACL-Anthology, and CiteSeer+DBLP data. This shows that the results depend considerably on the data set and on the pre-processing steps (e.g., whether PDF-to-text conversion is performed). 
  Overall, it can be assumed that the novel approaches to citation recommendation published in the near future will mainly be based on neural networks, too. 
 \end{enumerate}

 Overall, existing approaches are primarily based on implicit representations of the cited statements and concepts (e.g., embeddings of citation contexts \cite{Kobayashi2018,Han2018ACL}), but not on fine-grained explicit representations of statements or events. One reason for that might be the missing research on the different citation types besides the citation function, and the current relatively low performance of fact extraction and event extraction methods from text.
 \item \textbf{User model:}~As outlined in Sec.~\ref{sec:task-description-related-fields}, approaches to citation recommendation can optionally incorporate user information, such as the user name, the venue that the input text should be submitted to, or keywords which categorize the input text explicitly. Overall, we can observe that most approaches (12 out of 17, i.e., 71\%) do not use any user model. Five approaches are dependent on the author name of the citing document.\footnote{The two global citation recommendation approaches \cite{Liu2014JCDL,Bhagavatula2018} allow the user to disclose more information about her optionally.}
 \item \textbf{Prefilter:}~By default, all candidate papers need to be taken into account for any citation recommendation. This often results in millions of comparisons between representation forms and, thus, turns out to be unfeasible. To escape from that, the proposed methods often incorporate a pre-filtering step as a step before the actual recommendation, in which the set of candidate papers is drastically reduced. For instance, before applying a neural network-based approach for a precise citation recommendation, the top 2048 most relevant papers are retrieved from the paper collection via BM25 \cite{Ebensu2017}. In 30\% (5 out the 17) of the considered papers, the authors mention such a step (see Table~\ref{tab:comparison-approaches-part1}). 
 While three authors implement a certain numerical value as threshold \cite{Rokach2013,Livne2014SIGIR,Ebensu2017},\footnote{Examples in the case of global citation recommendation are \cite{Strohman2007SIGIR,Bhagavatula2018}.} others use flexible thresholds such as the word probabilities \cite{HuangWCMG15,Yin2017APWeb}.\footnote{Concerning global citation recommendation, we can refer here to \cite{Liu2014JCDL,Jiang2015CIKM}.} 
 \item \textbf{Citation context length:}~
 The size of the citation context varies from approach to approach. Typically, 1--3 sentences \cite{He2012SPIRE,Huang2012CIKM,Tang2014SIGIR,HuangWCMG15,Yin2017APWeb,Kobayashi2018} or a window of up to 50 words \cite{He2010WWW,Kataria2010,He2011WSDM,Rokach2013,Duma2014ACL,Livne2014SIGIR,Ebensu2017,Han2018ACL,Jeong2019arxiv} is used. 
 Investigations on the citation context length suggest that there is no one ideal citation context length~\cite{Alvarez2016}. 
 \item \textbf{Citation placeholders:}~
 The citation placeholders, i.e., the places in which a citation should be recommended, and therefore also the citation context, are typically already provided a priori for evaluating the single approaches (exceptions are \cite{He2011WSDM,Livne2014SIGIR}). The main reason for this fact is presumably that the past approaches focus on the citation recommendation task itself and see the identification of ``cite-worthy'' contexts as a separate task. Determining the cite-worthiness, which is similar to determining the citation function, is not tackled in the approaches. However, there have been separate attempts at solving this task \cite{FaerberECIR2018shortpaper,Sugiyama2010} (and related: \cite{Abu-JbaraR12}). 
 Also, with respect to performing the evaluation, having a flexible citation context makes it very tricky to compare the approaches in offline evaluations with the citation contexts and their citations from the ground truth. Single attempts such as \cite{He2011WSDM,Livne2014SIGIR}, solve it, however, for instance, by using only those citation contexts and associated citations which overlap with the found citation contexts to a considerable degree.
 \item \textbf{Cited papers' content needed: }
 The approaches to citation recommendation differ in the characteristic of whether they incorporate the content of the \textit{cited} documents or not. Incorporating the contents means that all cited documents need to be available in the form of full text. This is often a limitation, since any paper published somewhere could be referenced by authors; the cited documents are, thus, often not in any ready collection of citing documents. For instance, in the CiteSeer data set of \cite{Nallapati2008}, only 16\% of the cited documents are also citing documents; this is similar to the arXiv CS data set~\cite{Faerber2018LREC} and unarXiv data set \cite{Saier19}. Not incorporating the content, on the other hand, leads to a less fine-grained recommendation and the vision of even a single fact-based recommendation is illusive. 
 Considering the approaches to local citation recommendation, we cannot recognize a clear trend concerning the aspect of used content: both approaches using the cited papers' content and not using it have been proposed in recent years.
 \item \textbf{Evaluation data set:}~
 In general, a variety of data sets have been used in the publications. Most frequently (in 8 out of 17, i.e., 47\% of the cases), versions of the CiteSeer data set (i.e., CiteSeer, CiteSeerX, RefSeer) have been applied, because this data set has been available since the early years of citation recommendation research and because it is relatively large. However, even the approaches in recent years are often evaluated on newly created data sets. 
 As Sec.~\ref{sec:comparison-evaluation-data-sets} is dedicated to data sets used for citation recommendation, we can refer to this section for more details. 
 \item \textbf{Domain:}~
 Independent of which data set has been applied, all data sets cover the computer science or computational linguistics domain. We can assume that this is because (1) the papers in those domains are relatively easy to obtain online, and because (2) the papers are understandable by the authors of these approaches, allowing them to judge at first sight whether the recommendations are suitable. 
 \item \textbf{Evaluation metrics:}~
 Concerning the usage of evaluation metrics and the interpretation of evaluation scores, the following aspects are especially noteworthy: 
 \begin{enumerate}
 \item \textit{Varying metrics:} The metrics used across the papers vary considerably; most frequently, recall, MAP, nDCG, and MRR are used (10/9/7/7 out of 17 times). This variety makes it hard to compare the effectiveness of the approaches. 
 \item \textit{Varying data sets:} Since largely systems have been evaluated on varying data sets and with varying document filtering criteria, we can hardly compare the systems' performance overall. For instance, the recent approaches \cite{HuangWCMG15,Ebensu2017} report both nDCG@10 scores of 0.26.\footnote{In case of global citation recommendation, see \cite{Liu2016iConf} with an nDCG@10 score of 0.21.} 
 \item \textit{Varying $k$:} Even if the same metrics are used in different papers, and maybe when even the same data sets are used, for considering the top $k$ returned recommendations, different $k$ values are considered, with a great variance from $k=1$ up to $k=200$. Especially high values like $k=100$ \cite{Kobayashi2018} or $k=80$ \cite{Jeong2019arxiv} seem to be unrealistic as no user-friendly system would presumably expect the user to check so many recommendations. 
 \item \textit{Missing baselines:} 
 It can be observed that the considered papers do not reference all prior works (see also Fig.~\ref{fig:fig-cit})
 and that previously proposed approaches are not used sufficiently as baselines in the evaluations, although the papers propose solutions for the same research problem. 
 This applies to papers on local citation recommendation and global citation recommendation.  
 For instance, \cite{Jiang2018JCDL} does not cite \cite{Tang2014SIGIR}, although both tackle the cross-language citation recommendation problem. This issue was already observed for papers on paper recommendation in \cite{Beel2016}. 
 \item \textit{Varying citation recommendation tasks:} The system's performance strongly depends on the kind of citation recommendation which is pursued. Given not only a citation context as input, but also the metadata of the citing paper, such as the authors, the venue, etc., then the nDCG@10 score can be 0.62 as in \cite{Livne2014SIGIR} instead of around 0.26 as in \cite{HuangWCMG15,Ebensu2017}.\footnote{Moreover, global citation recommendation systems using only the papers' abstracts perform differently to the ones based on the papers' full text. This can be illustrated by the fact that Liu et al.~\cite{Liu2016iConf} use an abstract as input and obtain MAP@all of 0.16, while the same authors in \cite{Liu2014JCDL} obtain a MAP@all score of 0.64 when using the full text.}
 \end{enumerate}
\end{enumerate}
In total, it is very hard to compare the effectiveness of the approaches (1) if different metrics are used and with different top $k$ values, (2) if different evaluation data sets are used, (3) if the approaches do not use existing systems as baselines, and (4) if the differences in the task set-up are not outlined. Considering the above-discussed approaches, we can observe this phenomenon to a high degree.

\subsection{System Demonstrations} 

While a relatively large amount of approaches to citation recommendation have been published, only \textit{RefSeer} \cite{DBLP:conf/jcdl/HuangWMG14} and \textit{CITEWERTs}\footnote{\url{http://citewerts.citation-recommendation.org/}}~\cite{FaerberECIR2018demopaper} have been presented as systems for demonstration purposes. \textit{RefSeer} is based on the model proposed by He et al.~\cite{He2010WWW} and uses CiteULike as the underlying document corpus. It recommends one citation for each sentence in the input text. 
\textit{CITEWERTs}, in contrast, is the first system which not only recommends citations but also identifies cite-worthy contexts in the input text beforehand. 
This makes the system more user-friendly, since it hides unnecessary recommendations, and it reduces the number of costly recommendation computations. 
Besides these systems, to the best of our knowledge, only paper recommendation systems exist, i.e., systems that do not use any citation context, but, for instance, only use a citation graph~\cite{huynh2012scientific}. \textit{TheAdvisor}~\cite{DBLP:conf/jcdl/KucuktuncSKC13},  \textit{FairScholar}~\cite{DBLP:conf/ecir/Anand0D17} are further examples of paper recommender system demonstrations. \textit{Google Scholar},\footnote{\url{http://scholar.google.com/}} \textit{Mendeley}\footnote{\url{http://mendeley.com/}}, \textit{Docear}~\cite{Beel2011Docear}, and \textit{Mr. DLib}~\cite{Beel2011MrDLib} also provide a functionality for obtaining paper recommendations.

\section{Data Sets for Citation Recommendation} 
\label{sec:data-sets}

In this section, we give an overview of data sets which can be used in the context of citation recommendation. 
Section~\ref{sec:comparison-evaluation-data-sets} presents data sets containing papers' content, while Section~\ref{sec:working-data-sets} outlines data sets containing purely metadata about papers.

\subsection{Corpora Containing Papers' Content}
\label{sec:comparison-evaluation-data-sets}

\subsubsection{Overview of Data Sets} 

There exist several corpora which provide papers' content and, hence, can serve as a gold standard for automatic evaluations. Table~\ref{tab:data-sets-part1} gives an overview of the data sets which are considered by us.  
Note that we only consider data sets here that are not outdated and that are still available (either online or upon request from the author). Hence, old data sets, such as the Rexa data base~\cite{Strohman2007SIGIR} or the initial CiteSeer database~\cite{GilesBL98}, are not included.\footnote{\textit{CiteULike} (\url{http://www.citeulike.org/}), a popular data set for paper recommendation, is not included in our list, since the full text of the papers is not available.} 

Generally, we can differentiate between two corpora sets: firstly, the CiteSeer data sets, available in different versions, have been explicitly created for citation-based tasks. They already provide the citation contexts of each citing paper and can be described as follows: 
 \begin{itemize}
  \item \textbf{CiteSeerX (complete) \cite{Caragea2014}}: Referring to the CiteSeerX version of 2014, the number of indexed documents exceeded 2M. The CiteSeerX system crawls, indexes, and parses documents that are openly available on the Web. Therefore, only about half of all indexed documents are actually scientific publications, while a large fraction of the documents are manuscripts. The degree to which the findings resulting from the evaluations based on CiteSeerX also hold for the actual citing behavior in science is therefore unknown to some degree. 
  \item \textbf{CiteSeerX cleaned by Caragea et al. \cite{Caragea2014}}: The raw CiteSeerX data set contains a lot of noise and errors as outlined by Roy et al.~\cite{Roy2016}. Thus, in 2014, Caragea et al.~\cite{Caragea2014} released a smaller, cleaner version of it. The revised data set resolves some of the noise problems and in addition links papers to DBLP.
  \item \textbf{RefSeer \cite{HuangWCMG15}}: RefSeer has been used for evaluating several citation recommendation approaches~\cite{HuangWCMG15,Ebensu2017}. Since it contains the data of CiteSeerX as of October 2013 without further data quality improvement efforts, RefSeer is on the same quality level as CiteSeerX. 
  \item \textbf{CiteSeerX cleaned by Wu et al. \cite{Wu2017}}:  According to Wu et al.~\cite{Wu2017}, the cleaned data set \cite{Caragea2014} still has relatively low precision in terms of matching CiteSeerX papers with papers in DBLP. Hence, Wu et al. have published another approach for creating a cleaner data set out of the raw CiteSeerX data, achieving slightly better results on the matching of the papers from CiteSeerX and DBLP. 
 \end{itemize}

Then, there are collections of scientific publications, with and without provided metadata, for which citation contexts are not explicitly provided. However, in those cases, the citation contexts can be extracted by appropriate tools based on the papers' content, making these corpora also applicable as ground truth for offline evaluations. They are listed alphabetically in the following:
\begin{itemize}
 \item \textbf{ACL Anthology Network (ACL-AAN) \cite{Radev2013}}: ACL-AAN is a manually curated database of citations, collaborations, and summaries in the field of Computational Linguistics. It is based on 18k papers. The latest release is from 2016. ACL-AAN has been used as an evaluation data set for many tasks.
 \item \textbf{ACL Anthology Reference Corpus (ACL-ARC) \cite{Bird2008ACLARC}}:\footnote{\url{http://acl-arc.comp.nus.edu.sg/}.} ACL-ARC is a widely used corpus of scholarly publications about computational linguistics. There are different versions of it available. ACL-ARC is based on the ACL Anthology website and contains the source PDF files (about 11k for the February 2007 snapshot), the corresponding content as plaintext, and metadata of the documents taken either from the website or from the PDFs. 
 \item \textbf{arXiv CS} \cite{Faerber2018LREC}: This data set, used by \cite{FaerberECIR2018shortpaper,Faerber2019ECIR}, was obtained by utilizing all arXiv.org source data of the computer science domain and transforming the \LaTeX{} files into plaintext by an own implemented \TeX parser. As far as possible, each reference is linked to its DBLP entry.
 \item \textbf{CORE}:\footnote{\url{http://core.ac.uk/}.} CORE collects openly available scientific publications (originating from institutional repositories, subject repositories, and journal publishers) as data basis for approaches concerning search, text mining, and analytics. As of October 2019, the data set contains 136M open access articles. CORE has been proposed for citation-based tasks for several years. However, to the best of our knowledge, it has not yet been used for evaluating or deploying any of the published citation recommendation systems.
 \item \textbf{Scholarly Paper Recommendation Dataset 2 (Scholarly Data Set)}:\footnote{\url{http://www.comp.nus.edu.sg/~sugiyama/SchPaperRecData.html}.} This data set contains about 100k publications of the ACM Digital Library and has been used for evaluating paper recommendation approaches~\cite{Sugiyama2015,Sugiyama2013}. 
 \item \textbf{unarXiv} \cite{Saier19}: This data set is an extension of the \textit{arXiv CS} data set. It consists of over one million full text documents (about 269 million sentences) and links to 2.7 million unique papers via 29.2 million citation contexts (having 15.9 million unique references). All papers and citations are linked to the Microsoft Academic Graph. 
\end{itemize}

\subsubsection{Comparison of Evaluation Data Sets}

\begin{sidewaystable*}
\vspace{17.5cm}
\centering
\caption{Overview of data sets applicable to citation recommendation.}
\label{tab:data-sets-part1}
\begin{footnotesize}
\begin{tabular}{P{2.6cm}P{1.0cm}P{1.7cm}P{1.7cm}P{1.7cm}P{1.5cm}P{1.5cm}P{1.5cm}P{1.1cm}P{1.1cm}P{0.7cm}P{0.7cm}P{1.4cm}}
\toprule
& Size of data set & Citation context available, size & Metadata of citing paper (structured) & Metadata of cited paper (structured) & Full text of all citing papers & Full text of all cited papers & Abstract of citing paper & Abstract of cited paper & Full citation graph 
& Clean\-liness & Links & Usage \\
\toprule
CiteSeerX complete & very large & yes, 400~chars & yes (noisy) & yes (noisy) & yes & no & yes & not all & no (but large) & no &  no & \cite{Nallapati2008}\cite{Tang2009PAKDD} \cite{He2010WWW}\cite{Kataria2010}\cite{He2011WSDM} \cite{Huang2012CIKM}\cite{Rokach2013}\cite{HuangWCMG15}\\
\midrule
CiteSeerX cleaned by Caragea et al. & large & yes, 400~chars & yes (noisy) & yes (noisy) & no & no & yes & not all & no (but large) & no & DBLP & \\ 
\midrule
RefSeer & large & yes, 400~chars & yes (noisy) & yes (noisy) & no & no & yes & not all & no & no & no & \cite{Ebensu2017}\\
\midrule
CiteSeerX cleaned by Wang~et~al. & large & yes, 400~chars & yes (noisy) & yes (noisy) & no & no & yes & not all & no (but large) & no & DBLP & \\ 
\midrule
ACL-AAN & small & no (extractable) & yes & no (extractable) & yes (noisy) & no & (extractable) & not all & no & no & no & \cite{Duma2014ACL}\cite{Han2018ACL} \cite{Yang2018IEEEAccess}\cite{Jeong2019arxiv}\\
\midrule
ACL-ARC & small & no (extractable) & yes & no (extractable) & yes (noisy) & no & (extractable) & not all & no & no & no & \cite{Bethard2010}\\
\midrule
arXiv CS & medium & yes, 1~sentence & yes & yes & yes & no & (extractable) & not all & no & yes & DBLP &  \\
\midrule
CORE & very large & no (part. extractable) & yes & no & partially & no & yes & not all & no (but large) & yes & no & \\
\midrule
Scholarly Dataset~2 & medium & no (extractable) & no (extractable) & no (extractable) & yes & no &  (extractable) & not all & no & yes & DBLP &  \\
\midrule
unarXiv & large & yes, 3~sentences & yes & yes & yes & no & (extractable) & not all & no & yes & MAG & \\
\bottomrule
\end{tabular}
\end{footnotesize}
\vspace{2\baselineskip} 
\caption{Overview of papers' metadata data sets applicable to citation recommendation.}
\label{tab:data-sets-part2}
\begin{footnotesize}
\begin{tabular}{P{3.9cm}P{2.2cm}P{2.2cm}P{2.2cm}P{2.2cm}P{1.7cm}P{3.1cm}}
\toprule
 & Size of data set & Abstract of citing paper & Abstract of cited paper & Full citation graph & Clean\-liness & Links \\
\toprule
AMiner DBLPv10 & large & partially & partially & yes & yes & DBLP \\ \midrule
AMiner ACMv9 & large & yes & yes & yes & yes & DBLP (but no URIs) \\ \midrule
Microsoft Academic Graph & very large & no & no & yes & yes & no \\ \midrule
Open Academic Graph & very large & yes & yes & yes (open access papers) & yes & DBLP (but no URIs) \\ \midrule 
PubMed & large & no & partially & yes & yes & no \\
\bottomrule
\end{tabular}
\end{footnotesize}
\end{sidewaystable*}

Table~\ref{tab:data-sets-part1} shows the mentioned data sets categorized by different dimensions. We can outline the following highlights with respect to these dimensions:

\paragraph{Size of data set} 
The considered data sets differ considerably in their sizes: they range from small (below 100k documents; see ACL-ARC and ACL-AAN) to very large (over 1M documents; see CiteSeerX complete). Note thereby that the cleanliness of the provided papers' contents does not necessarily depend on the overall size of the data set: for instance, ACL-AAN and ACL-ARC are quite noisy, as they contain rather old publications, which are hard to parse. However, clean metadata of the cited papers is available for those data sets.

\paragraph{Availability of citation context}
CiteSeerX, arXiv CS, and the unarXiv data set provide explicitly extracted citation contexts of the citations in the documents. In case of the different versions of CiteSeerX, a fixed window of 400 characters has been chosen around the citation markers. In the case of arXiv CS and unarXiv, the content is provided sentence-wise, so that all sentences annotated with citation identifiers can be used as citation context. 
The corpora which contain the publications contents in their original form -- namely, ACL-AAN, ACL-ARC, CORE, and Scholarly -- do not provide citation contexts. However, these contexts could be extracted without much effort by using appropriate tools from the source PDF files.

\paragraph{Structured metadata of citing papers}
For all the presented corpora, structured metadata of all the citing papers is provided. An exception is Scholarly, which only consists of PDF files. Hence, the metadata needs to be extracted by oneself with the corresponding tools. 
Note that the metadata is clean only for those corpora for which the information has been entered manually at some point. 
For CiteSeerX, all information, including the metadata of citing papers, has been extracted from the publications (mainly PDFs). Hence, this framework is independent of external data. However, as a tradeoff, the extracted metadata is to some extent noisy and inaccurate (missing information or wrongly split strings etc.)~\cite{Roy2016}.

\paragraph{Structured metadata of cited papers}
Only the CiteSeer data sets as well as arXiv CS and unarXiv provide this information per se. 
In the case of CORE, it is planned that publications will be linked to the Microsoft Academic Graph.
Consequently, structured metadata of cited papers will be retrievable from this data set.\footnote{As of November 4, 2019, the webpage mentions links to the Microsoft Academic Graph. However, no corresponding information can be found in the data set.} 
For the other corpora containing publications' content, the metadata of the cited papers can be obtained by extracting the information from the publications' reference sections via the appropriate tools. However, note that it does not only require the parsing via an appropriate information extraction tool, but also the reconciliation of the data (i.e., building a global database of publications' metadata). 
The task of how to find out if two referenced papers are actually the same and, hence, should have the same identifier is non-trivial and is known as \textit{citation matching}.

\paragraph{Paper content of citing papers}
Some approaches, such as sequence-to-sequence approaches, require the complete contents of all citing papers. 
In the complete CiteSeerX data set, all citing papers' contents are still available.
Also the paper collections Scholarly, arXiv CS, unarXiv, ACL-ARC, and ACL-AAN (and CORE to some degree) contain the papers' full texts. 
However, in case of Scholarly and ACL-AAN, the original data sets do not contain the contents as plaintext, so that one first needs to run appropriate transformation approaches. 

\paragraph{Paper content of cited papers} 
All considered data sets do not provide the full texts of all cited papers. This is not surprising, as papers typically cite papers without any restrictions and, thus, from various publishers.

\paragraph{Abstract of citing papers}
Since the abstract of papers belongs to the metadata, it is quite easily obtainable for both citing papers and cited papers. Furthermore, it already summarizes the main points of each paper (although typically not sufficiently for a detailed and precise recommendation) and can be used for obtaining a better representation of the paper, and, hence, for improving the recommendation of papers based on citation contexts overall. 
Regarding the citing papers, all data sets either provide the abstract already in an explicitly given form (see the CiteSeerX data set and partially CORE) or contain the original publications (as PDF or similar formats), so that the abstract can be extracted from them (see Scholarly, arXiv CS, unarXiv, ACL-ARC, ACL-AAN).

\paragraph{Abstract of cited papers}
Having as much information as possible about what the cited papers are dealing with is crucial for a good citation recommendation. In this context, the abstracts of cited papers are very useful and are used by several approaches \cite{He2010WWW,He2011WSDM,Livne2014SIGIR,Jiang2015CIKM,Lu2011CIKM,Duma2014ACL,Yin2017APWeb,Bhagavatula2018,Jiang2018JCDL}. 
However, none of the data sets contain abstracts for all cited papers. 

\paragraph{Full citation graph} 
In a full citation graph (also called \textit{citation network}), not only the citations of the citing papers are represented, but the citations of any paper of a given document collection. 
Such a graph can be used for obtaining a good representation of the papers (see paper embeddings \cite{Ebensu2017,Gipp2014}) and to compute similarities among papers. 
None of the considered corpora provides such an extended citation graph.\footnote{Note, however, that data sets such as unarXiv and CORE 
link to the Microsoft Academic Graph providing citation information.}
As an alternative, one can think of linking papers from one corpus with papers of a metadata corpus (see Section~\ref{sec:working-data-sets}).

\paragraph{Cleanliness} 
The situation is mixed in this regard: the metadata of the papers is of good quality, especially if it originates from corresponding, dedicated databases instead of being extracted solely from the publications themselves (see ACL-AAN, ACL-ARC, arXiv CS, and unarXiv vs. the CiteSeerX data sets). 
The papers' content is typically provided via information extraction methods, meaning that the quality is not that high, particularly if the papers were hard to parse and process, e.g., due to being very old (see the papers of ACL-ARC and ACL-AAN vs. Scholarly, which contains newer papers) or due to special formating in the publications, such as formulas in the text (see CiteSeerX data sets vs. the arxiv CS and unarXiv data sets, where formulas were detected and removed).

\paragraph{Links to bibliographic data sets}
Having publications linked to external bibliographic data sets such as DBLP allows the use of interlinked information for paper representations and for search.
Corpora of scientific papers have often been created in the area of computer science, since there are many publications available online. As a consequence, the most widely used bibliographic database for computer science, DBLP, has been used as a reference of interlinking. More precisely, the cleaned versions of CiteSeerX and the arXiv CS data set provide links to DBLP. unarXiv provides links to the Microsoft Academic Graph, as it covers not only computer science papers, but also many other disciplines.

\subsection{Corpora Containing Papers' Metadata}
\label{sec:working-data-sets}

Besides corpora including papers' content, data sets exist that contain metadata about publications; typical metadata include the citation relations between papers and the titles, venues, publication years, and abstracts of the publications. 
Although no content is usually provided, the metadata can be regarded as an explicit, structured representation of the papers and, hence, can be used as a valuable representation of the papers, e.g., for learning embedding vectors based of them (see, e.g., \cite{Ebensu2017,Ganguly2017}). 
Due to their extensive sizes, the following data sets are in our view particularly suitable for citation recommendation:\footnote{The data set \textit{Mendeley DataTEL} is not listed, as it has not been available to us after several requests. Further data sets, such as CORA (\url{https://relational.fit.cvut.cz/dataset/CORA}), have not been shortlisted due to their small size. We have also not listed bibliographic databases like DBLP here, as they contain neither the papers' contents nor information about the citations between papers. Also Springer's SciGraph does not contain any citation information yet. Bibliographic databases, such as \textit{Scopus} and \textit{Web of Science}, are dedicated information retrieval platforms, but do not officially support bulk downloads.}

\begin{itemize}
 \item \textbf{AMiner DBLPv10}\footnote{\url{https://aminer.org/citation}.} \cite{Tang2008}: This data set contains over 3M papers and 25.2M citation relationships, making it a large citation network data set. Since DBLP was used as data source, the data is very clean.
 \item \textbf{AMiner ACMv9}\footnote{\url{https://aminer.org/citation}.} \cite{Tang2008}: This data set has the same structure as AMiner DBLPv10, but was constructed from 2.4M ACM publications, with 9.7M citations.
 \item \textbf{Microsoft Academic Graph}:\footnote{\url{https://www.microsoft.com/en-us/research/project/microsoft-academic-graph/}.} This data set can be considered as an actual knowledge graph about publications and associated entities such as authors, institutions, journals, and fields of study. Direct access to the MAG is only provided via an API. However, dump versions have been created.\footnote{\url{https://kddcup2016.azurewebsites.net/} and \url{http://ma-graph.org/}.} Prior versions of the MAG are known as the \textit{Microsoft Academic Search data set}, based on a the project \textit{Microsoft Academic Search} which retired in 2012.
 \item \textbf{Open Academic Graph}:\footnote{\url{https://www.openacademic.ai/oag/}.} This data set is designated to be an intersection of the Microsoft Academic Graph and the AMiner data. In many cases, the DBLP entries for computer science publications ought to be retrievable.
 \item \textbf{PubMed}:\footnote{\url{https://www.nlm.nih.gov/databases/download/pubmed_medline.html.}} PubMed is a database of bibliographic information with a focus on life science literature. As of October 2019, it contains 29M citations and abstracts. It also provides links to the full-text articles and third-party websites if available (but no content).
\end{itemize}

Table~\ref{tab:data-sets-part2} shows the mentioned data sets categorized by various dimensions. 
The same dimensions are used as for comparing the corpora in Section~\ref{sec:comparison-evaluation-data-sets}, except 
the ones which are homogeneous among the metadata data sets (e.g. \textit{availability of citation context, paper content of citing papers}). 
Due to page limitations, we omit a textual comparison of the mentioned metadata data sets.

\section{Evaluation Methods and Challenges}
\label{sec:evaluation-methods-challenges}

In this section, we first discuss the different ways of evaluating citation recommendation approaches. 
Secondly, we point out important challenges related to evaluating citation recommendation approaches. 
Afterwards, we provide the reader with guidelines concerning what aspects to consider for evaluating future recommender systems.

\subsection{Evaluation Methods for Citation Recommendation}
\label{sec:evaluation-methods-citation-recommendation}

Generally, we can distinguish between offline evaluations, online evaluations, and user studies. In offline evaluations, no users are involved and the evaluation is performed automatically. Online evaluations measure the acceptance rates of recommendations in deployed recommender systems. 
User studies are used for measuring the user satisfaction through explicit user ratings.

For \textit{offline evaluations}, the following evaluation methods have been applied so far for citation recommendation:
\begin{enumerate}
 \item \textbf{Strict ``citation re-prediction:''}~
 This evaluation method has been used by almost all approaches 
 to local citation recommendation (15 out of 17; see \cite{He2010WWW,Kataria2010,He2012SPIRE,Huang2012CIKM,Rokach2013,Liu2013AIRS,Duma2014ACL,Tang2014SIGIR,HuangWCMG15,Yin2017APWeb,Ebensu2017,Kobayashi2018,Han2018ACL,Yang2018IEEEAccess,Jeong2019arxiv}). 
 \\
 The evaluation is performed as follows: an approach is evaluated by assessing which of the citations that have been recommended by the system are also in the original publications. We can therefore call this method ``re-prediction.'' This evaluation method scales very well, but ignores several evaluation challenges, such as the relevance of alternative citations, and the cite-worthiness of contexts (see Section~\ref{sec:evaluation-challenges}). 
 Hence, the evaluation metrics used for strict citation re-prediction, such as normalized discounted cumulative gain (nDCG), mean average precision (MAP) and mean reciprocal rank (MRR), might reflect the reality in the sense of the citing behavior observed in the past, but not the desired citing behavior. 
 \item \textbf{``Relaxed citation re-prediction:''}~
 In order to allow papers to be recommended which are not written as citations by the authors of the papers, but which are still relevant, and on the other hand, to keep the evaluation still automatic and scalable, sometimes a relaxation of the strict re-prediction method is applied. In the set of considered approaches, the following methods have been applied by both He et al.~\cite{He2011WSDM} and Livne et al.~\cite{Livne2014SIGIR}: 
 \begin{enumerate}
   \item The \textit{relative co-cited probability} metric is designed as a modified accuracy metric and based on the assumption that papers which are frequently co-cited are relevant to each other. Hence, if not the actual cited paper, but a co-cited paper\footnote{$B$ is a co-cited paper of $A$, if both $A$ and $B$ are cited by a third paper $C$.} is recommended, this paper is also considered as a hit to some degree.
   The relative co-cited probability is the ratio to which recommended papers are either directly cited or are co-citations of actual citations. In the latter case, the co-cited paper is only scored gradually. 
   \item The regular nDCG score is used for measuring the correct ranking of items. Modifying this score is based on the idea that if the actual paper is not standing on the intended position, but there is another paper there, which is also relevant (here, again determined by the co-citations), then this should also be judged as correct to some degree. More specifically, the authors use the average relative co-cited probability of $r$ with all original citations of $d$ to obtain a citation relevance score of $r$ to $d$. Then the documents in $D$ are sorted with respect to this relevance score and each document is assigned a score on a 5 point scale regarding its relevance. Finally, the average nDCG score over all documents is calculated based on these scores. 
 \end{enumerate}
\end{enumerate}

A more comprehensive, but not very scalable way to evaluate approaches is to rely on \textit{online evaluations} \cite{Beel2015Comparison}.
None of the considered approaches has been evaluated in this way so far. Also no \textit{user studies} for citation recommendation systems are known to us.\footnote{For paper recommendation, a few manual evaluations exist \cite{Beel2016}. However, paper recommendation is out of our scope.}

\subsection{Challenges of Evaluating Citation Recommendation Approaches}
\label{sec:evaluation-challenges}

In the previous subsection we learned that it is hard to apply traditional evaluation metrics for offline evaluations of citation recommendation systems. We now point out further challenges when it comes to determining the performance of citation recommendation systems. 
In Section~\ref{sec:guidelines}, we then propose steps for approaching some of these challenges.

\subsubsection{Fitness of Citations} 
\label{sec:fitness-citations}

Training and evaluating a citation recommendation system based on an existing paper collection used as ground truth is tricky, since the citing behavior encoded in the citations of these considered papers is taken as ground truth. This becomes a problem when the original citing behavior is not favorable and adaptations are desired. 
In the past, several analyses of scientific citing behavior have been published \cite{Tahamtan2018}. We can reuse these for characterizing the different aspects of citing biases in the context of evaluating citation recommendation. We thereby group citing biases along the attributes of the \textit{citable} publications: 
\begin{enumerate}
 \item \textbf{Content Understanding: } 
 Authors of citing papers may differ in their expertise, knowledge level, and working style when selecting citations (cf. professor vs. masters student).
 The suitability of the content of citable papers is therefore often judged differently. 
 
 Furthermore, authors of citing papers might perform literature investigations and reviews in a rather sloppy way \cite{Ishita2018} and read, for instance, mainly titles and abstracts of documents only. However, titles and abstracts may deceive users about the true claims and contributions of papers. 
 Moreover, the selection of citations can be biased by the style of the titles and abstracts (see, e.g., \cite{ButerR11,SuboticM14}). 
 Also the writing style of the fulltext of the citable papers has some influence on citing, as it reflects the perceived quality of the paper~\cite{Liu97}. 
 \item \textbf{Authors:} 
 It is quite common to cite publications written by oneself, called self-citations, \cite{Aksnes03,Hyland2003} or written by colleagues, advisors, and friends \cite{WangS98}, with an element of preferential bias. Although analyses have shown that this is not per se harmful \cite{Tahamtan2018}, a citation recommender system ought to be designed independent of any bias. Furthermore, the user of a citation recommendation system might be interested particularly in works she does not yet know.
 
 There are also cases in which the authors of the citing and cited document do not know each other, but in which the author of the citing document still favors specific persons as authors of the cited documents. Most notably, sometimes citation cartels exist in the scientific communities, which first of all cite papers within sub-communities \cite{Fister2016TowardTD}. Furthermore, it has been observed that even the country a person comes from, the race, and the gender play a role in the selection of citations \cite{TahamtanAA16}. A bias towards citing authors who act as the referee or reviewer of the citing document in a peer-review process is also plausible~\cite{WangW99}. 
 \item \textbf{Venue and Paper Type:}
 It is obvious that the venue is an influential factor in selecting appropriate citations for a given text. Highly rated conferences and journals might get higher levels of attention and are privileged compared to lower rated conferences, workshops, and similar publication formats \cite{Callaham2002,Ware2015STM}. A bias can go so far that a relatively weak publication in a prestigious journal receives a high number of citations only due to the centrality of the journal \cite{Callaham2002}. Papers in interdisciplinary journals are more likely to be cited \cite{Annalingam2014}. 
 Last but not least, it should be noted that, in the frame of the widely performed peer-reviewing process, especially papers that were published in the same venue as the citing paper are more often selected as citations \cite{Wilhite2012}. 
 
 Many venues have introduced a page limit for submitted papers. As a consequence, authors often choose to cut several citations which would be relevant and important for understanding the content. 
 \item \textbf{Timeliness:} 
 The temporal dimension concerning citing behavior is, to the best of our knowledge, relatively unexplored in the context of citation recommendation. On the one hand, due to the acceleration in the publishing rate of scientific contributions, authors of citing papers might target citing especially recent papers. On the other hand, older papers have more citations and are easier to find. 
 Note also that the reasons for citing specific publications can change over time \cite{CaseH00}. 
 \item \textbf{Accessibility and Visibility: }
 During the citing process, researchers are limited by their capabilities for finding appropriate publications for citing. In particular, they typically cite only papers to which they have fulltext access. However, a considerable amount of researchers have limitations in this regard, such as having no license for accessing papers of specific publishers (e.g., ACM or Springer) and paper collections. 
 Consequently, the set of citable papers is narrowed down considerably. Hence, either not all concepts and claims in the citing paper can be backed up by citations or they cannot be backed up by optimal citations. 
 
 Papers are also embedded in the social interactions and dissemination processes of researchers.
 Most notably, the claim that prominent publications get cited more is comprehensible and well-studied, even though more relevant alternative publications might exist for citation~\cite{White2004AL}. Prominent papers are papers which already have a high number of citations, or papers written by authors who are well known in the field and who also have a high aggregated citation count. 
 We can refer in this context to the studies on the so-called \textit{Matthew effect}~\cite{Beel2009Google} and on the \textit{Google Scholar effect}~\cite{Serenko2015}. 
 Particularly prominent papers are called \textit{landmark papers} and \textit{citation classics} \cite{Small04}. They are characterized by the fact that they are often added as citations in a ritualized way and self-enforce their citing. 
 
 Last but not least, it cannot be neglected that nowadays many publications are disseminated via social networks and other channels. Research on these aspects in the context of citing behavior has been performed only to a limited extent \cite{Lin2013}.
 \item \textbf{Discipline:}
 Firstly, researchers naturally work within scientific communities and disciplines, with the consequence that they are often exclusively familiar with works published in their discipline or field and that it is difficult for them to discover papers from other fields (due to different venues, terminology, etc.). Hence, citations tend to be limited by the affiliation to the discipline (or even research field). 
 
 Secondly, the citing behavior also changes from discipline to discipline. Comparing the citing behavior across disciplines, and, hence, comparing also citation recommendation systems trained and tested on different disciplines, is challenging. 
 For instance, disciplines differ in (1)~the number of articles published, (2)~the number of co-authors per paper, (3)~the relevance of the publication type (e.g., journal, conference, book) for publishing, and (4)~the age of cited papers \cite{Mabe2011}. These aspects have a direct influence on the relevance function of any citation recommendation model. Investigations and evaluations on the context of citation recommendation approaches are missing so far, however. As stated in Section~\ref{sec:comparison-approaches}, evaluations on citation recommendation have been performed mainly on corpora containing only computer science publications.
\end{enumerate}

\subsubsection{Cite-Worthiness of Contexts}

Citation recommendation systems typically consider predefined citation contexts for their prediction. However, in reality, typically not only the provided citation contexts are cite-worthy, but also further contexts. Among others, one reason for missing citations is the page restriction which authors need to fulfill for submitting papers to venues.\footnote{The San Francisco Declaration on Research Assessment (DORA; \url{http://www.ascb.org/dora/}) from 2012 targets the improvement of ways in which the outputs of scientific research are evaluated, and was signed by over 13,000 researchers and institutions. In this declaration, it is proposed that authors should not be restricted by page limitations for references any more, or at least should have reduced restrictions. The reality, however, still looks different.} 
In the past, there have been a few approaches for assessing the cite-worthiness of potential citation contexts automatically, however, only in the sense of a binary classification task~\cite{FaerberECIR2018shortpaper,Bonab2018,Sugiyama2010,FaerberECIR2018demopaper}. 
Although there are single works on characterizing the citation context, such as on the citation function, the citation importance, and the citation polarity (see Section~\ref{sec:related-research-fields}), these aspects are not considered in citation recommendation approaches so far. In particular the type of citation, given as the citation function or in the form of another classification, such as whether the citation backs up a single concept or a claim, seems to be a notable aspect to be considered.

\subsubsection{Scenario Specificity}

As outlined in Section~\ref{sec:scenarios-citation-recommendation}, citation recommendation systems can be applied in different scenarios, differing in particular in (1) the user type (see expert vs. non-expert setting) and (2) in the type and length of input text.
Considering these nuances during evaluation makes a comparison of approaches difficult. However, it is necessary, as the comparison would be unfair otherwise. For instance, citation recommendation systems using only text from an abstract perform differently than ones based on a paper's full text (see the MAP@all score of 0.16 \cite{Liu2016iConf} vs. 0.64 \cite{Liu2014JCDL}). 
In contrast to that, the difference in the usability of systems for different user types can be assessed via online evaluations and user studies.

\subsection{Discussion}
\label{sec:guidelines}

Based on the given observations, we propose the following suggestions for an improved evaluation of citation recommendation systems: 

\paragraph{Concerning \textbf{offline evaluations}}
In the main, nDCG, MRR, MAP, and recall have been used as the evaluation metric in existing offline evaluations. We recommend using them for the top $k$ recommendations with a rather low value for $k$ (e.g., $k=5$ or $k=10$) as in \cite{Huang2012CIKM,Livne2014SIGIR,HuangWCMG15,Han2018ACL}, since it is in our view realistic to return only very few recommendations to the user per citation context (and not using e.g., nDCG@50, and nDCG@75 as in \cite{He2010WWW} or MAP@100 as in \cite{Tang2014SIGIR}). 
Tang and Zhang~\cite{Tang2009PAKDD} agree with us that it is hard to specify for each citation context how many recommended citations should be returned and notes that for simplicity, the average number of citations per paper could be set as $k$ (e.g., 11 in \cite{Tang2009PAKDD}), if the whole input document is considered. 
Common evaluation metrics used for citation recommendation reflect the reality only in the sense of the citing behavior observed in the past, but not alternatively valid citations.  
So far, only a few citation recommendation systems have been evaluated based on alternative offline evaluation metrics (see ``relaxed citation re-prediction'' in Section~\ref{sec:evaluation-methods-citation-recommendation}). For instance, the precision metric is softened and papers are also assessed as a hit if they are only related to the cited publications in the citation graph. We argue that such metrics need to be taken with care in the light that citation recommendation aims to back up specific claims and concepts. 

\paragraph{Concerning \textbf{online evaluations and user studies}}
As outlined in Section \ref{sec:evaluation-methods-citation-recommendation}, user studies and online evaluations are so far missing in the context of citation recommendation, while offline evaluations predominate. The situation is therefore similar to the situation in the field of paper recommendation~\cite{Beel2016}.  Similar to \cite{Beel2015Comparison}, we recommend performing user studies and online evaluations as necessary steps in the future. 
This might be particularly fruitful (1) for determining a reasonable ratio of citations per document (cf. cite-worthiness of contexts), and (2) for assessing the relevance of alternative citations, which can be even more relevant than the original citations.\footnote{The fact that other documents are more relevant as citations can also be observed for Wikipedia, see \cite{FetahuMNA16}.}  
Differentiating and automatically determining different levels of relevance seems to be necessary to address this issue, as outlined by \cite{Strohman2007TechReport}. 
Studies on the importance and grading of citations are rare (see Section~\ref{sec:related-citation-based-tasks}), and, to the best of our knowledge, there are no user assessment studies on assessing alternative papers in the context of (personalized or unpersonalized) citation recommendation.

\paragraph{Concerning \textbf{citing biases}}
In order to minimize the biases in the citing behavior, the corpora used for training and testing might need to be changed. For instance, only those publications might be considered for which a high degree of fairness can be guaranteed. 
Single publications could be classified in this respect and might receive a confidence value concerning biases \cite{Pitoura2017}.

To not introduce a citing bias via recommending specific papers, citation recommendation systems should use large paper collections (see Sec.~\ref{sec:data-sets}) and the information which recommendation algorithm and candidate papers are used, should be made available to the user. 

\paragraph{Concerning \textbf{scenario specificity}}
Similar to paper recommender systems~\cite{Beel2016}, the evaluation results of citation recommendation approaches are often not reproducible, since the data sets are not available and/or many important details of the implementation are omitted in the papers due to constraints such as page limitations \cite{BeelBLLG16}. Therefore, we recommend making evaluation data sets, the implementation of the system, and the calculation of evaluation metrics as transparent as possible. Also the targeted scenario (see Section~\ref{sec:scenarios-citation-recommendation}) and use case characteristics should be clearly visible.

\section{Potential Future Work}
\label{sec:potential-future-work}

There are still many variations of the architectures and of the input and output of citation recommendation systems which have not been considered yet. More specifically, we can think of the following adaptations to enhance and improve citation recommendation:
 \begin{itemize}
 \item Topically diversifying recommended citations \cite{Chakraborty2015ICDE};
 \item Recommending papers which state similar, related, or contrary claims as the ones in the citation contexts (i.e., recommending not only papers with identical claims);
 \item Inserting a sufficient (optimal) set of citations; this could be useful in the presence of paper size limitation, which may be imposed, for example, by  conferences. A citation recommendation system should then prioritize important citation contexts that cannot be left without the insertion of citations, while perhaps skipping other less important ones in order to keep the paper size within the limits;
 \item Given an input text with already present citations, suggesting newer/better ones to update some obsolete/poor citations;
 \item Combating the cold-start problem for freshly published papers which are not yet cited, hence no training data is available on them;
 \item Incorporating information on social networks among researchers and considering knowledge sharing platforms; such data can offer additional (often timely) hints on the appropriateness of papers to be cited in particular citation contexts;
 \item Focusing on specific user groups, which have a given pre-knowledge in common (see our listed scenarios in Sec. \ref{sec:scenarios-citation-recommendation}); 
 \item Studying the influences of citing behavior on citation recommendation systems and developing methods for minimizing citing biases in citation recommendation such as biases arising from researchers belonging to the same domains, research groups, or geographical areas (cf. Section~\ref{sec:evaluation-challenges});
 \item Developing global context-aware citation recommendation approaches, i.e., approaches that recommend citations in a context-aware way, yet still consider the entire content of a paper;
 \item Recommending citations refuting an argument (using argumentation mining);
 \item Designing domain-specific citation recommendation approaches and evaluating generic approaches on different disciplines (outside computer science).
\end{itemize}

Besides these concrete future works, we can think of the following visions in the long term, which embrace a new process of citing in the future: 
\begin{enumerate}
 \item One can envision that, in the future, citation recommendation approaches could better capture the semantics of the citation context, with the result that actual fact-based citation recommendation would have good chance to become reality. This suggests the opportunity of obtaining precise citation recommendations, since both the claims in the citation context and the claims in the candidate cited documents are represented explicitly in a semantically-structured form. In this sense, citation recommendation systems might be capable of not only citing publications, but also any knowledge (in particular, facts and events) available on the Web. This vision becomes particularly feasible in light of the Linked Open Data (LOD) cloud and is in line with research on LOD-based recommender systems \cite{Noia2012}.
 \item One can envision that the working style of researchers would dramatically change in the next few decades \cite{Casati2007liquid,Montuschi2014}. As a result, we might think not only of citation recommendation as considered in this article, but one based on the expected or potential characteristics of scientific publishing. For instance, one can imagine that publications will not be published in PDF format any more, but in either an annotated and more structured version of it (with information about the hypotheses, the methods, the data sets, the evaluation set-up, and the evaluation results), or in the form of a flexible publication format (e.g., subversioning system), in which authors can subsequently change the content, especially the citations, since over the time citations might become obsolete or new citations might become relevant. 
\end{enumerate}

\section{Conclusions and Outlook}
\label{sec:conclusions}

In this survey, we gave a profound overview of the research field of citation recommendation. To that end, we firstly introduced citation recommendation via outlining possible scenarios and via a description of the task. 
We saw that the approaches to context-aware citation recommendation can be grouped into hand-crafted feature-based models, topic models, machine translation models, and neural network models. The approaches do not only differ with respect to the underlying method, but also with respect to the provided input data. More specifically, the considered set-ups differ in the use of a user model, the prefiltering of candidate papers, the length of the citation context, whether citation placeholders are provided, and whether the content of cited papers is needed. Concerning the evaluation, the approaches are evaluated based on very diverse metrics and different data sets, making it hard to assess the validity and advance of single approaches. Moreover, approaches are often compared to existing approaches to a limited extent. 

We also considered the data sets that can be used for deploying and evaluating citation recommendation. We distinguished between corpora containing papers' content and corpora providing papers' metadata. Here we learned that several corpora exist, especially in the field of computer science. However, the data sets differ considerably in their size and in their quality (e.g., noise due to information extraction). 

Concerning the challenges of evaluating citation recommendation and the evaluation methods used so far, we found out that biases in the citing behavior have largely been ignored, as well as the ``worthiness'' to cite at all or in specific circumstances. Assessing citation recommendations might also depend on the scientific discipline and on the concrete use case. Approaches have been evaluated rather unilaterally and not across disciplines. 

Upcoming approaches on citation recommendation are likely to be based on more advanced techniques of machine learning, such as variants of recurrent neural networks. 
In the long term, one can envision that citation recommendation approaches can better capture the semantics of the citation context, with the result that actual fact-based citation recommendation becomes reality. 
Given the likely continuation and proliferation of the ``tsunami'' of publications and of citations in the years and decades to come, we can assume that citation recommendation will become an integral component of a researcher's working environment.

\bibliographystyle{plain}

\end{document}